\documentclass[aps,twocolumn,pra,superscriptaddress,tightenlines]{revtex4-1}
\usepackage{amsfonts}
\usepackage{graphicx}
\usepackage{hyperref}
\usepackage{amsmath}
\usepackage{amssymb}
\usepackage[usenames, dvipsnames]{color}
\usepackage{subfigure}
\usepackage{lipsum}

\begin{document}

\title{Quadrupole Effects on Nuclear Magnetic Resonance Gyroscopes}
\author{Yue Chang}
\email{yuechang7@gmail.com}
\author{Shunag-Ai Wan}
\author{Jie Qin}
\affiliation{Beijing Automation Control Equipment Institute, Beijing 100074,
China}
\affiliation{Quantum Technology R$\&$D Center of China Aerospace
Science and Industry Corporation, Beijing 100074, China}



\begin{abstract}
Nuclear magnetic resonance gyroscopes that detect rotation as a shift in the
precession frequency of nuclear spins, have attracted a lot of attentions.
Under a feedback-generated drive, the precession frequency is supposed to be
dependent only on the angular momentum and an applied magnetic field. However,
nuclei with spins larger than 1/2, experience electric quadrupole interaction with electric field gradients at
the cell walls. This quadrupole interaction shifts the precession
frequencies of the nuclear spins, which brings inaccuracy to the rotation
measurement as the quadrupole interaction constant $C_q$ is difficult to
precisely measure. In this work, the effects of quadrupole interaction on
nuclear magnetic resonance gyroscopes is theoretically studied. We find that,
when the constant $C_q$ is small compared to the characteristic decay rate
of the system, as the strength of the feedback-driving field increases, the
quadrupole shift monotonically decreases, regardless of the sign of $C_q$.
In large $C_q$ regime, more than one precession frequency exists, and the
nuclear spins may precess with a single frequency or multi-frequencies
depending on initial conditions. In this regime, with large driving
amplitudes, the nuclear spin can restore the single-frequency-precession.
These results are obtained by solving an effective master equation of the
nuclear spins in a rotating frame, from which both the steady-state
solutions and dynamics of the system are shown.
\end{abstract}

\maketitle

\section{Introduction}

\begin{figure}[tbp]
\begin{center}
\includegraphics[width=0.85\linewidth]{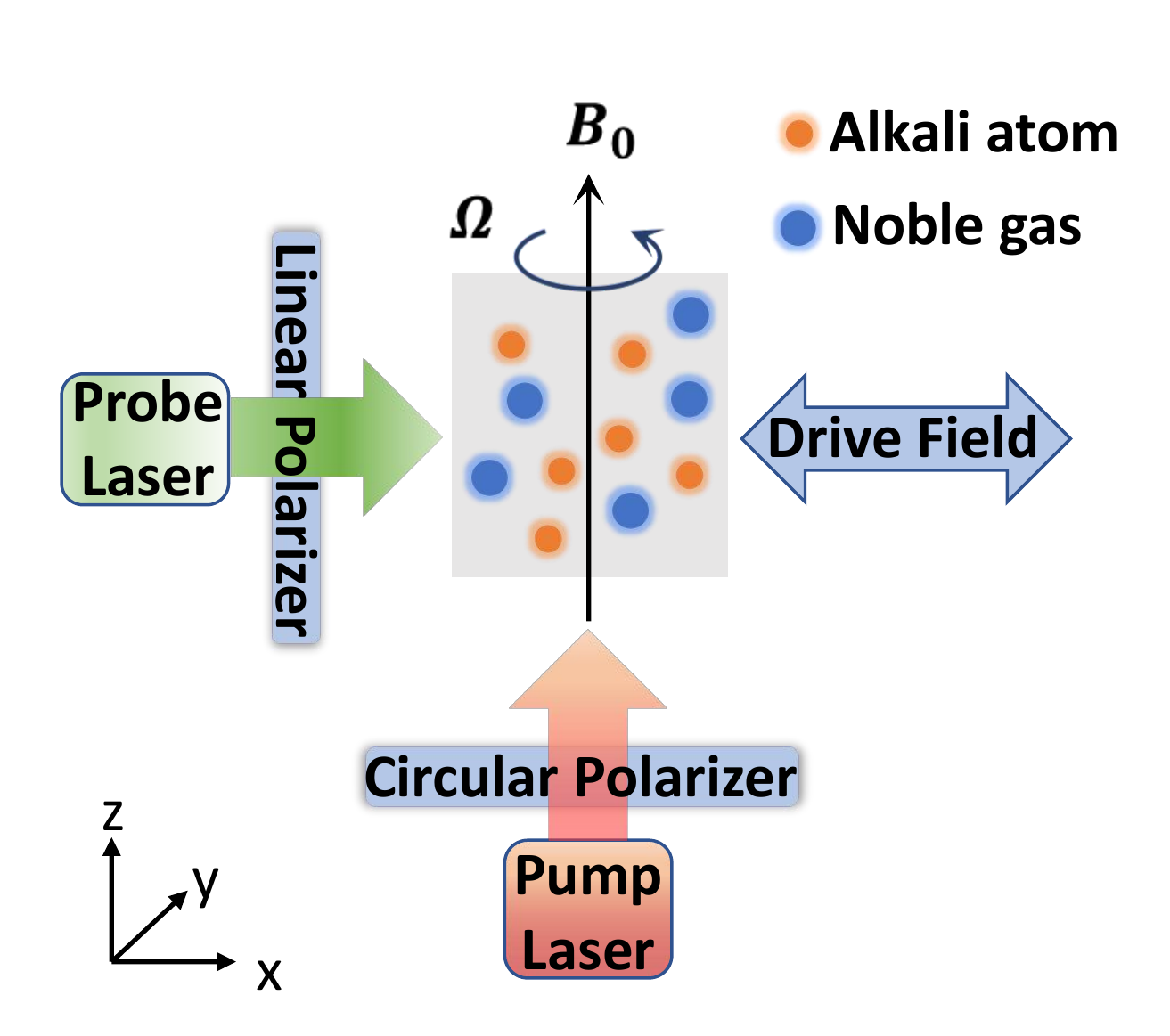}
\end{center}
\caption{Schematic of a NMRG, where alkali atoms (orange circles) and noble
gas (blue circles) are confined in an atomic cell (usually there is also
buffer gas that is not shown in the figure) which is subjected to a rotation
and an applied static magnetic field $\vec{B}_{0}$\ along the $z$ direction.
The alkali atoms are optically pumped by a circularly polarized laser,
resulting in polarization of the electron spins. This electronic
polarization is transferred to nuclear spins of the noble gas by
spin-exchange collisions. In presence of a feedback-driven AC magnetic field in
the $x$ direction, the polarized nuclear spins precess with a frequency that
depending on the rotation and the applied field $\vec{B}_{0}$. This
precession frequency is acquired by measuring the optical rotation using a
linearly-polarized laser with the in-situ alkali-metal-vapor atomic
magnetometer.}
\label{fig1}
\end{figure}

Nuclear magnetic resonance gyroscopes (NMRGs) accomplish rotation detection by
measuring a shift in the nuclear spins' precession frequency in an applied
magnetic field \cite%
{simpson1963optically,bayley1973optically,walker2016spin,donley2010nuclear,walker2015synchronously,PhysRevLett.95.230801,popov2019theoretical}%
. As an inertial navigation system, a NMRG has advantages such as high precision
and compact size. Thus, it has attracted much attentions in several decades.
A NMRG with basic elements is schematically shown in Fig.~\ref{fig1},
where a rotating atomic vapor cell (with its angular momentum along the $z$
direction) contains alkali-metal atoms and noble gas. Usually there is some buffer
gas (not shown in Fig.~\ref{fig1}) to reduce radiation trapping \cite%
{molisch1998radiation} and wall-collisions of the alkali atoms. A static
magnetic filed $\vec{B}_{0}$\ along the $z$ direction is applied in this
atomic cell, and a circularly polarized laser is used to optically pump \cite%
{happer2010optically,auzinsh2010optically,RevModPhys.44.169,PhysRevA.99.063411}
the alkali atoms. Then the nobles gas are polarized via spin exchange \cite%
{PhysRevA.29.3092,walker1997spin,kornack2002dynamics,jau2002high,ma2011collisional}
with the polarized alkali atoms, and in presence of a feedback-generated AC
magnetic field along the $x$ axis, the noble-gas nuclear spins precess with
a frequency that depends on the applied magnetic field and rotation of the
system. This precession frequency is then measured by the in-situ
alkali-metal-vapor atomic magnetometer \cite%
{Budker2007,PhysRevLett.93.160801,PhysRevLett.94.123001,walker1997spin},
where the alkali atoms's polarization is probed by a linearly polarized
laser.

However, the nuclear precession frequency of the noble gas can be shifted by
quadrupole interactions \cite{slichter2013principles}. It is given by
Wigner-Eckart theorem \cite{sakurai1995modern} that all nuclei with spins
larger than $1/2$ have quadrupole moments. For instance, to stabilize the
magnetic field, dual-isotope NMRGs are used \cite%
{grover1979nuclear,walker2016spin}, where one of these two isotopes in the commonly used schemes
has its nuclear spin larger than $1/2$. This quadrupole interaction \cite%
{kwon1981quadrupole,PhysRevLett.59.1480,wu1988coherent,heimann1981quadrupole,heimann1981quadrupole2,donley2009nuclear}%
, resulting from collisions of the noble gas atoms with the cell walls,
brings inaccuracy to the rotation measurement since the quadrupole
interaction constant is difficult to be precisely measured. A phase delay $\beta
$ (with respect to the precession of the nuclear spins) in the feedback
drive can suppress this frequency shift \cite{walker2016spin}. It is shown
that the precession frequency shift is a monotonic function of $\beta $,
where the frequency shift is zero at a certain value $\beta _{0}$. However, $%
\beta _{0}$ is strongly dependent on the system parameters such as the
longitudinal and transverse relaxation rates of the nuclear spins, the sign
and strength of the quadrupole interaction constant $C_{Q}$, the amplitude
of the feedback drive $\Omega _{d}$, and so on. Thus, to have a reduced
frequency shift by tuning $\beta $, one needs knowledge of other parameters
of the system, especially when the system parameters fluctuate, one needs to
know the time dependence of these parameters. It makes this method less
practical.

In literature, there are many other researches on the frequency shift
\cite%
{kwon1981quadrupole,PhysRevLett.59.1480,wu1988coherent,heimann1981quadrupole}
and nuclear spin relaxation \cite%
{kwon1981quadrupole,wu1988coherent,heimann1981quadrupole2} induced by the electric quadrupole interaction, but few are on
feedback-driven systems \cite{walker2016spin}. Therefore, in this paper, we
theoretically study the effect of the quadrupole interaction on
feedback-driven NMRGs. By solving an effective master equation of the
nuclear spins, we find that the absolute value of the quadrupole shift
monotonically decreases as the feedback-driving amplitude $\Omega _{d}$
increases, despite other parameters of the system. This can be understood
intuitively: the quadrupole interaction brings linearity in the nuclear
spin's energy-level spacings, which will shift the resonant frequency in the
long-term limit. But increasing the driving amplitude will reduce the
relative nonlinearity by reducing the ratio between the difference of the
energy-level spacing and the corresponding energy-level spacing. As a
result, the frequency shift is reduced. Apart from the frequency shift, we
show that the the quadrupole interaction reduces the nuclear precession
magnitude because of the destructive interference of oscillations between
adjacent states of the nuclear spins. In principle, this reduction can be
compensated by increasing the density of the noble gas while keeping a relatively high sensitivity of the in-situ atomic
magnetometer.

For large quadrupole interaction, more than one precession frequency exists.
In this case, the nuclear spin precesses with one or multi-frequencies
depending on its initial condition. We study the dynamics of the nuclear
spins for different initial states. We show that increasing $\Omega _{d}$ can stimulate
more oscillations with different frequencies. However, when the driving
amplitude $\Omega _{d}$ is sufficiently large, the nuclear spin restores the
single-frequency precession. The mechanism is the same as in the small-$%
\Omega _{d}$ regime: large driving amplitude can reduce the nonlinearity
coming from the quadrupole interaction.

The rest of the paper is organized as follows. In Sec.~\ref{s2}, we give the
effective master equation for the noble gas' nuclear spins, which is
obtained by neglecting correlations between the alkali atoms and noble gas
atoms and eliminating the alkali atoms' degrees of freedom. In Sec.~\ref{s3}%
, we write the master equation in a rotating frame with respect to the
feedback driving frequency that needs to be determined. The long-term
behaviors of the precession frequency are shown from symmetry analysis of
this master equation with a time-independent Liouvillian. Then in In Sec.~%
\ref{s4}, we numerically solve this master equation and show the
steady-state behaviors as functions of the driving amplitude $\Omega _{d}$.
The absolute value of the frequency shift decreases monotonically as $\Omega
_{d}$ increases. For comparison, we also show the quadrupole shift as a
function of$\ $the phase shift\ $\beta $. As the quadrupole interaction
constant $C_{Q}$ increases, the system may exhibit more than one precession
frequencies. This situation is studied in Sec.~\ref{s5}, where we show
dynamics of the nuclear spins for various initial conditions and driving
amplitude $\Omega _{d}$. Finally, in Sec.~\ref{s6}, we summarize our results.

\section{Effective master equation for nuclear spins with nonzero quadrupole
moments}

\label{s2}

In this section, we will give the effective master equation of the noble-gas
nuclei. The master equation constitutes a unitary-evolution part which is
describe by a system Hamiltonian $H$ and dissipations $\mathcal{L}_{D}\left[
\rho \right] $ \cite{Gardiner2004,Walls2008}:%
\begin{equation}
\partial _{t}\rho =-i\left[ H,\rho \right] +\mathcal{L}_{D}\left[ \rho %
\right] ,  \label{1}
\end{equation}%
where $\rho $ is the density matrix of the noble-gas's nuclear spin. The
Hamiltonian contains three terms%
\begin{equation}
H=\omega _{0}K_{z}+C_{Q}K_{z}^{2}+\Omega _{d}K_{x}\sin \left( \phi -\beta
\right) ,
\end{equation}%
where $K_{j}$ with $j=x,y,z$, is the nuclear spin operator of the noble gas.
The first term in $H$ is the interaction of the nuclear spins to constant
applied fields: an applied magnetic field $B_{0}\hat{e}_{z}$ and rotation of
the gyroscope apparatus along the $z$ direction with frequency $\Omega $
(see Fig.~\ref{fig1}). Thus, the \textquotedblleft ideal\textquotedblright\
precession frequency $\omega _{0}$ of the noble gas' nuclear spin in the
laboratory frame is%
\begin{equation}
\omega _{0}=\gamma B_{0}\mp \Omega ,
\end{equation}%
where $\gamma $ is the gyromagnetic ratio and the sign \textquotedblleft $%
\mp $\textquotedblright\ depicts that the angular momentum is in the $\pm z$
direction. The second term in $H$ is the quadrupole interaction to
collision-induced electric gradients at the cell walls. It is this
quadrupole interaction which results in deviation of the nuclear spin's
actual precession frequency from the ideal one $\omega _{0}$. Here, we have
assumed the atomic cell is a $z$-axial cylinder, or a $z$-axial right
regular prism excluding a cube. Averaging over the sticking time and the
entire inner surface of the vapor cell, one can acquire the quadrupole
interaction Hamiltonian simply as $C_{Q}K_{z}^{2}$ \cite{PhysRevLett.59.1480}%
, where a constant that depends on the nuclear spin $K$ has been neglected.
To drive the nuclear spins to precess at its resonant frequency $\omega _{0}$%
, a feedback-generated ac magnetic field in the $x$-direction is applied.
The amplitude $\Omega _{d}$ of this oscillating field is a constant, but the
phase $\phi $ is feedback generated via the phase of $\left\langle
K_{+}\right\rangle \equiv \left\langle K_{x}+iK_{y}\right\rangle $, i.e., $%
\phi $ is defined through%
\begin{equation}
\left\langle K_{+}\right\rangle \equiv K_{\perp }e^{-i\phi },
\end{equation}%
where $K_{\perp }$ is the absolute value of $\left\langle K_{+}\right\rangle
$.\ Here, a constant phase delay $\beta $ is employed, which can suppress
the quadrupole shift \cite{walker2016spin}. For small driving amplitude $%
\Omega _{d}$ so that the actual precession frequency is much larger than $%
\Omega _{d}$, we can utilize the rotating-wave approximation to $H$ as%
\begin{equation}
H=\omega _{0}K_{z}+C_{Q}K_{z}^{2}+\frac{\Omega _{d}}{4i}\left(
K_{+}e^{i\left( \phi -\beta \right) }-K_{-}e^{-i\left( \phi -\beta \right)
}\right) ,
\end{equation}%
where we have neglected the fast-oscillating terms that oscillate with twice
of the precession frequency.

To acquire enhanced precession signals of the nuclear spins from collective
effects, nuclear spins are polarized through spin-exchange interaction
(Fermi-contact hyperfine interaction) with alkali atoms \cite%
{PhysRevA.29.3092,walker1997spin,jau2002high,ma2011collisional}:%
\begin{equation}
H_{se}=\alpha \left( R\right) \vec{K}\cdot \vec{S},
\end{equation}%
where $\vec{S}$ is the electron spin operator of the alkali atom, $\alpha
\left( R\right) $ is the coupling coefficient depending on the internuclear
separation $R$ between the alkali atom and noble gas atom. It is known that
this Fermi-contact hyperfine interaction transfers electron spins to nuclear
spins in two ways \cite{PhysRevA.29.3092,walker1997spin}: through binary
collisions between a pair of an alkali atom and a noble gas atom, and though
three-body-collision formed weakly-bound van der Waals molecules in presence
of the buffer gas. Tracing out the alkali atom's degrees of freedom by
neglecting correlations between alkali atoms and noble gas nuclei, we have
spin-exchange induced dissipation as \cite%
{PhysRevA.58.1412}%
\begin{eqnarray}
&&\mathcal{L}_{se}\left[ \rho \right]  \notag \\
&=&\Gamma _{S}\left[ \vec{K}\cdot \rho \vec{K}-\frac{1}{2}\left\{ \rho
,K^{2}\right\} \right]  \notag \\
&&+2\Gamma _{f}\left\langle F_{z}^{2}\right\rangle \left[ 2K_{z}\rho
K_{z}-\left\{ \rho ,K_{z}^{2}\right\} \right]  \notag \\
&&+\Gamma _{f}\left\langle F^{2}-F_{z}^{2}\right\rangle \left[ K_{+}\rho
K_{-}+K_{-}\rho K_{+}-\left\{ \rho ,K^{2}-K_{z}^{2}\right\} \right]  \notag
\\
&&+\Gamma _{p}\left[ K_{+}\rho K_{-}-K_{-}\rho K_{+}+\left\{ \rho
,K_{z}\right\} \right] ,
\end{eqnarray}%
where $\Gamma _{S}=\Gamma _{bin}+\Gamma _{s}$ is the total S-damping rate,
with the binary collision rate $\Gamma _{bin}$, $\Gamma _{s}/\Gamma _{f}$\
is the exchange rate originating from the short/long-lived van der Waals
molecules, and $\Gamma _{p}=\Gamma _{S}\left\langle S_{z}\right\rangle
+\Gamma _{f}\left\langle F_{z}\right\rangle $ is the polarization rate.
Here, $F=I+S$, where $I$ is the nuclear spin of the alkali atom.
Consequently, taking into account the nuclear spin's longitudinal and
transverse relaxation due to other mechanisms, we have the dissipation term
in the master equation (\ref{1}) as%
\begin{eqnarray}
&&\mathcal{L}_{D}\left[ \rho \right]  \notag \\
&=&\Gamma _{1}\left[ \vec{K}\cdot \rho \vec{K}-\frac{1}{2}\left\{ \rho
,K^{2}\right\} \right] +\Gamma _{2}\left[ 2K_{z}\rho K_{z}-\left\{ \rho
,K_{z}^{2}\right\} \right]  \notag \\
&&+\Gamma _{p}\left[ K_{+}\rho K_{-}-K_{-}\rho K_{+}+\left\{ \rho
,K_{z}\right\} \right] ,  \label{2}
\end{eqnarray}%
where $\Gamma _{1}=\Gamma _{1}^{\prime }+\Gamma _{S}+2\Gamma
_{f}\left\langle F^{2}-F_{z}^{2}\right\rangle $ and $\Gamma _{2}=\Gamma
_{2}^{\prime }+2\Gamma _{f}\left\langle 2F_{z}^{2}-F^{2}\right\rangle $,
with $\Gamma _{1,2}^{\prime }$ the relaxation rate from wall collisions,
magnetic field gradients, and other possible mechanisms rather than the
spin-exchange interaction.

We note that in the master equation (\ref{1}), the effective magnetic field
from the electron magnetization has been ignored since it is very small
compared to $\omega _{0}$. And because of the much slower dynamics of the
nuclear spins compared to the typical time scale of the the alkali atoms' $z$%
-directional polarization, $\Gamma _{1,2,p}$ in Eq.~(\ref{2}) can be treated
as constants. In spite of these simplification, Eq.~(\ref{1}) is still a
nonlinear equation whose exact solution can only be obtained numerically.
Before we do the numerics, in Sec.~\ref{s3}, we will show some properties of
the actual nuclear precession frequency in the long-term limit.

\section{Master equation in a rotating frame}

\label{s3}

In this section, we will write Eq.~(\ref{1}) in a rotating frame with a
time-independent Liouvillian. For this purpose, we assume the long-time
behavior of $\phi $ can be written as $\phi =-\left( \omega _{0}-\omega
\right) t+\theta _{0}$, where $\omega $ is the deviation of the actual
precession frequency from the ideal one, and $\theta _{0}$ is a constant
phase. This assumption is valid when the quadrupole interaction constant $%
C_{Q}$\ is small compared to the characteristic decay rate of the system.\
Note that the system has $U(1)$ symmetry, i.e., the master equation (\ref{1}%
) is invariant under the transformation $\rho \rightarrow e^{iK_{z}\theta
}\rho e^{-iK_{z}\theta }$, and $K_{+}\rightarrow K_{+}e^{i\theta }$. As a
result, the steady state has a phase uncertainty, which is determined by
initial conditions. Since we only concern the precession frequency, for
simplicity, $\theta _{0}$ is set to be zero here. Then, in a rotating frame
with $\tilde{\rho}=e^{iK_{z}\left( \omega _{0}-\omega \right) t}\rho
e^{-iK_{z}\left( \omega _{0}-\omega \right) t}$, the master equation can be
rewritten as%
\begin{equation}
\partial _{t}\tilde{\rho}=-i\left[ \tilde{H},\tilde{\rho}\right] +\mathcal{L}%
_{D}\left[ \tilde{\rho}\right] .  \label{3}
\end{equation}%
where the Hamiltonian (with rotating-wave approximation) in the rotating
frame is%
\begin{equation}
\tilde{H}=\omega K_{z}+C_{Q}K_{z}^{2}+\frac{\Omega _{d}}{4i}\left(
K_{+}e^{-i\beta }-K_{-}e^{i\beta }\right) .
\end{equation}%
This master equation with the time-independent Liouvillian has a
steady-state solution $\tilde{\rho}_{s}$\ which is a function of $\omega $
that needs to be determined. Back to the original frame, the average value
of $K_{+}$ in the long-term limit is%
\begin{equation}
\left\langle K_{+}\right\rangle =\text{Tr}\left( K_{+}\tilde{\rho}%
_{s}\right) e^{i\left( \omega _{0}-\omega \right) t}.
\end{equation}%
As a result, $\omega _{s}$, the solution of $\omega $ is determined via the
condition%
\begin{equation}
\text{angle}\left[ \text{Tr}\left( K_{+}\tilde{\rho}_{s}\right) \right] =0,
\end{equation}%
i.e.,%
\begin{equation}
k_{y}(\omega _{s},\Omega _{d},C_{Q},\beta )\equiv \mathrm{Tr}\left( K_{y}%
\tilde{\rho}_{s}\right) =0  \label{7}
\end{equation}%
and%
\begin{equation}
k_{x}(\omega _{s},\Omega _{d},C_{Q},\beta )\equiv \mathrm{Tr}\left( K_{x}%
\tilde{\rho}_{s}\right) >0.
\end{equation}

Performing the time-reversal transformation $\mathcal{T}$ that gives $%
\mathcal{T}K_{y}\mathcal{T}^{-1}=-K_{y}$ and $\mathcal{T}i\mathcal{T}%
^{-1}=-i $, one can acquire that the dissipation $\mathcal{L}_{D}$ remains
unchanged, but the Hamiltonian $\tilde{H}$ becomes%
\begin{equation*}
\mathcal{T}\tilde{H}\mathcal{T}^{-1}=\omega K_{z}+C_{Q}K_{z}^{2}-\frac{%
\Omega _{d}}{4i}\left( K_{+}e^{i\beta }-K_{-}e^{-i\beta }\right) .
\end{equation*}%
Taking into account of the sign change of $i$ in the master equation, we have%
\begin{equation}
k_{x}(\omega ,\Omega _{d},C_{Q},\beta )=k_{x}(-\omega ,\Omega
_{d},-C_{Q},-\beta )  \label{4}
\end{equation}%
and%
\begin{equation}
k_{y}(\omega ,\Omega _{d},C_{Q},\beta )=-k_{y}(-\omega ,\Omega
_{d},-C_{Q},-\beta ).  \label{5}
\end{equation}%
Eq.~(\ref{5}) shows that in absence of the quadrupole interaction, $%
k_{y}(\omega =0)=0$ when the phase delay $\beta =0$. In this case, $\omega
=0 $ may be a solution, depending on whether $k_{x}(\omega ,\Omega
_{d},C_{Q},\beta )>0$. To further study it, we perform a unitary transformation%
\begin{equation}
U=\exp \left( i\pi K_{z}\right) ,
\end{equation}%
which is actually a $\pi $-rotation along the $z$-axis, to the master
equation \ref{3}. This transforms $K_{x}\rightarrow -K_{x}$ and $%
K_{y}\rightarrow -K_{y}$. So we obtain%
\begin{equation}
k_{x,y}(\omega ,\Omega _{d},C_{Q},\beta )=-k_{x,y}(\omega ,-\Omega
_{d},C_{Q},\beta ).  \label{6}
\end{equation}%
Namely, when the sign of $\Omega _{d}$ is changed, the sign of $k_{x,y}$ is
also changed. Considering that the solution $\omega _{s}$ requires positive $%
k_{x}(\omega _{s},\Omega _{d},C_{Q},\beta )$, we can conclude that in absence of the quadrupole interaction, either
positive or negative $\Omega _{d}$ enables the nuclear spin to precess with
the ideal frequency $\omega _{0}$.

In presence of the quadrupole interaction, the frequency deviation $\omega
_{s}$ is nonzero. One way to suppress it is to employ the phase shift $\beta
$, as shown in Ref.~\cite{walker2016spin}. However, to determine $\beta $,
one needs the knowledge of $C_{Q}$. In fact, it can also be seen from Eq.~(%
\ref{5}) that when the sign of $C_{Q}$ changes, the sign of $\beta $ should
also be changed to insure $k_{y}(\omega _{s}=0)=0$. Since $C_{Q}$ is
difficult to measure accurately, we propose another solution to reduce the
quadrupole shift: in next section, we will show that increasing the driving
strength $\left\vert \Omega _{d}\right\vert $ will monotonically reduce
the quadrupole shift $\left\vert \omega _{s}\right\vert $, despite other
system parameters.

\section{Reduction of the quadrupole shift}

\label{s4}

\begin{figure}[tbp]
\begin{center}
\includegraphics[width=0.48\linewidth]{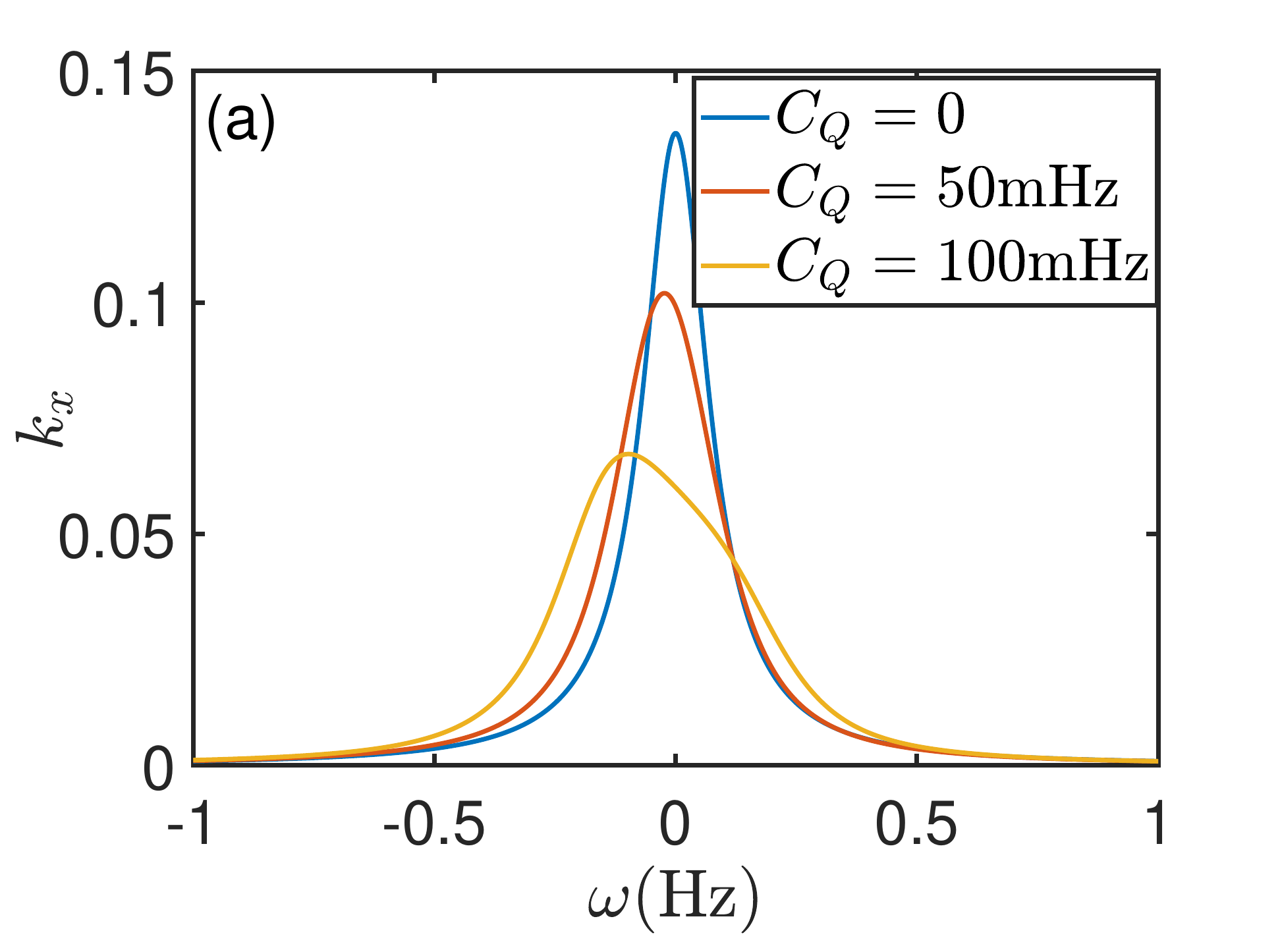} \includegraphics[width=0.48%
\linewidth]{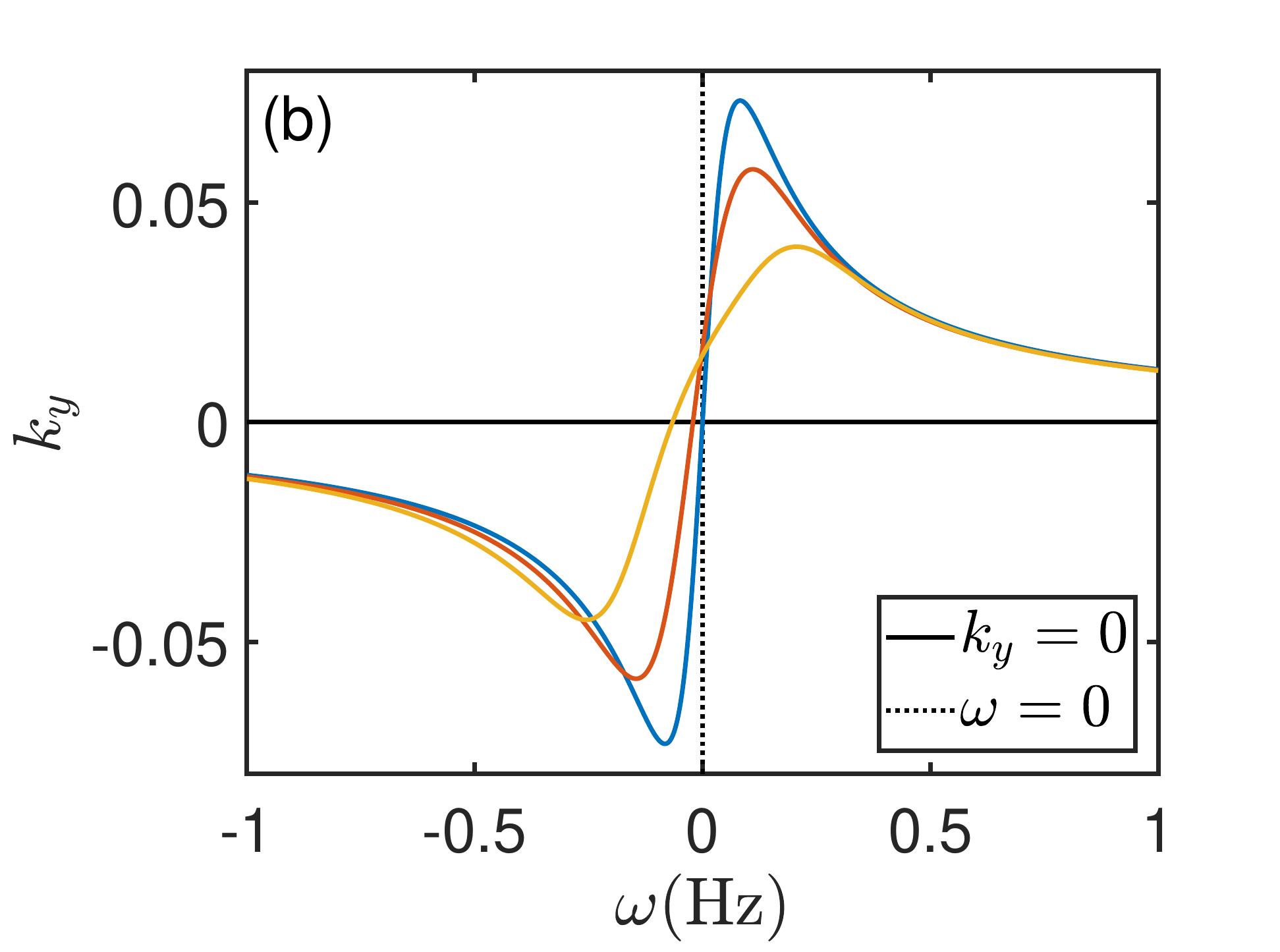}
\end{center}
\caption{Average values (a) $k_{x}$ and (b) $k_{y}$ as functions of $\protect%
\omega $ with different quadrupole interaction strength $C_{Q}$ for the the
spin-$3/2$ case. Other parameters are: $\Gamma _{1}=\Gamma _{2}=1/20$~Hz, $%
\Omega _{d}=1/20$~Hz, $\protect\beta =0$. With positive $\Omega _{d}$, $%
k_{x} $ is always positive, as shown in (a). Thus, negative $\Omega _{d}$
will not give a nontrivial solution of $\protect\omega $ since the solution requires
that $k_{x}>0$, and Eq.~(\protect\ref{6}) gives that $k_{x}<0$ for negative $%
\Omega _{d}$. In (b), the intersection of $k_{y}$ and the $k_{y}=0$ line
(solid black line) gives the solution $\protect\omega _{s}$, and the dotted
black line shows the $\protect\omega =0$ case. It is shown that with
positive $C_{Q}$, the frequency deviation is negative, while Eq.~(\protect
\ref{5}) gives that the frequency deviation is positive for negative $C_{Q}$%
. When $C_{Q}$ increases, the absolute value of the deviation also
increases. }
\label{fig2}
\end{figure}

In this section, we will numerically solve the master equation \ref{3} and
acquire the precession-frequency shift $\omega _{s}$. We first need to
determine whether positive or negative $\Omega _{d}$ gives zero frequency
deviation in absence of the phase delay $\beta $ and quadrupole interaction.
Without loss of generality, we consider a nucleus with spin 3/2, which is the
nuclear spin of $\left. ^{131}\mathrm{Xe}\right. $ that is commonly used in
dual-isotope NMRGs. With other parameters $\Gamma _{1}=\Gamma _{2}=1/20$~Hz,
$\Omega _{d}=1/20$~Hz, $\beta =0$, we plot $k_{x,y}$ as functions of $\omega $
for different quadrupole interaction constants $C_{Q}$ in Fig.~\ref{fig2}.
Fig.~\ref{fig2}(a) shows that $k_{x}$ is positive with positive $\Omega _{d}$%
. Thus, from the requirement that $k_{x}(\omega _{s},\Omega _{d},C_{Q},\beta
)$ must be positive and Eq.~(\ref{6}), we conclude that negative $\Omega
_{d} $ does not give nontrivial solutions of $\omega $. Thus, in the rest of the paper, we
will consider positive $\Omega _{d}$ only. With positive $k_{x}$,\ $\omega
_{s}$ is determined by $k_{y}\left( \omega \right) =0$, which is the
intersection of $k_{y}$ with the solid black line in Fig.~\ref{fig2}(b). It
is shown that when $C_{Q}>0$, the actual precession frequency deviate from
the ideal one $\omega _{0}$. This deviation becomes larger when $C_{Q}$
increases.

\begin{figure}[tbp]
\begin{center}
\includegraphics[width=0.48\linewidth]{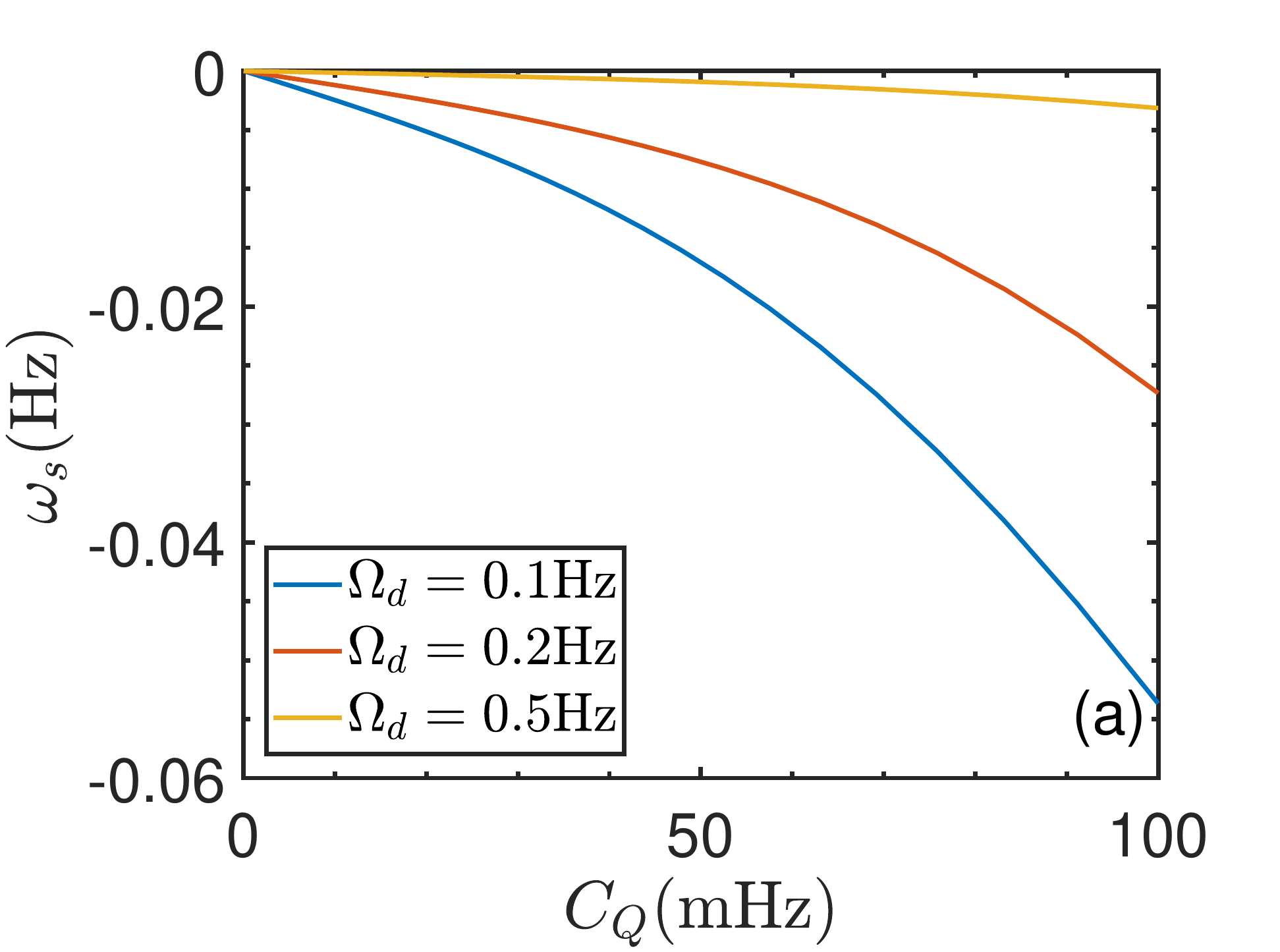} %
\includegraphics[width=0.48\linewidth]{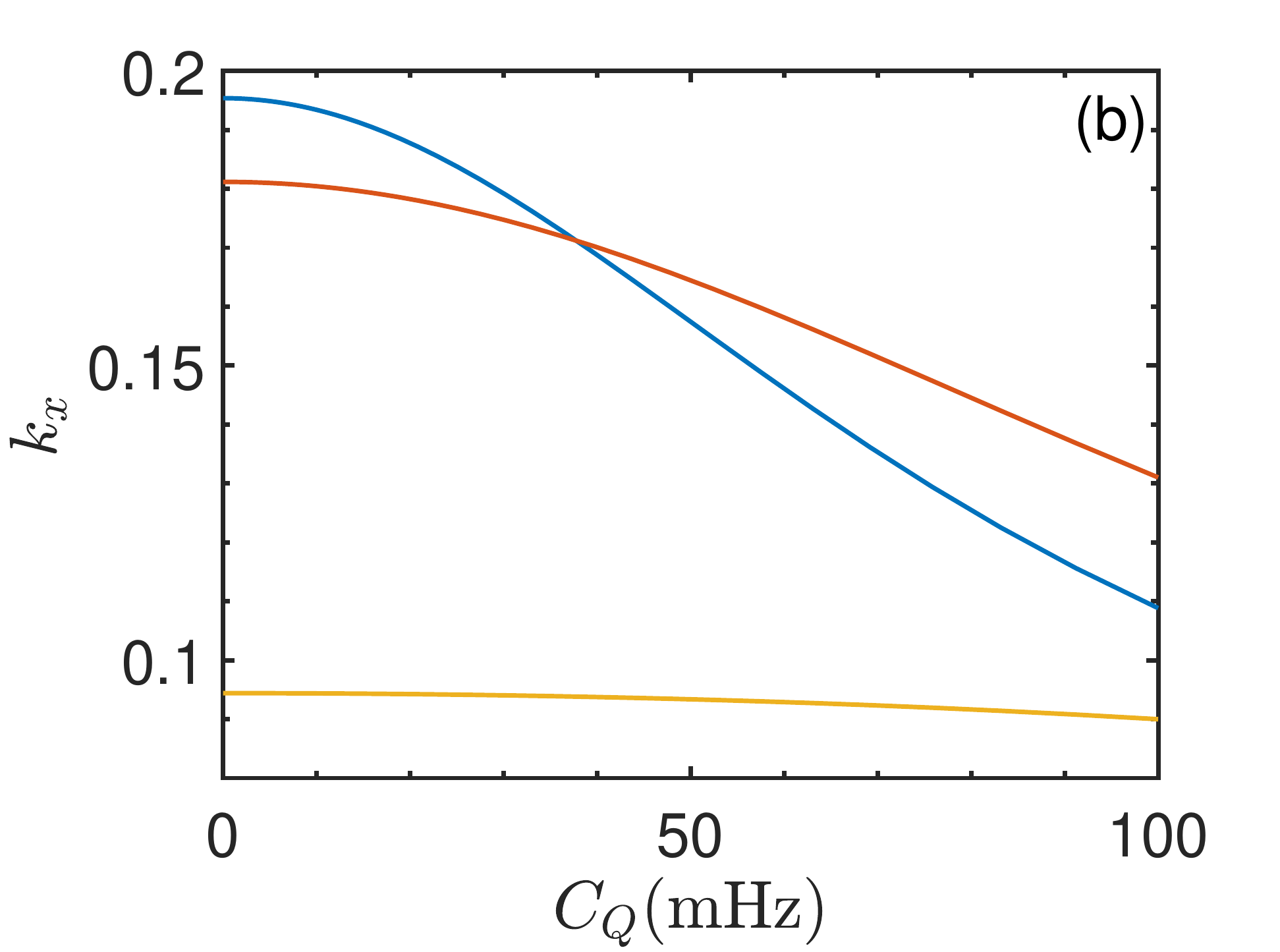} %
\includegraphics[width=0.48\linewidth]{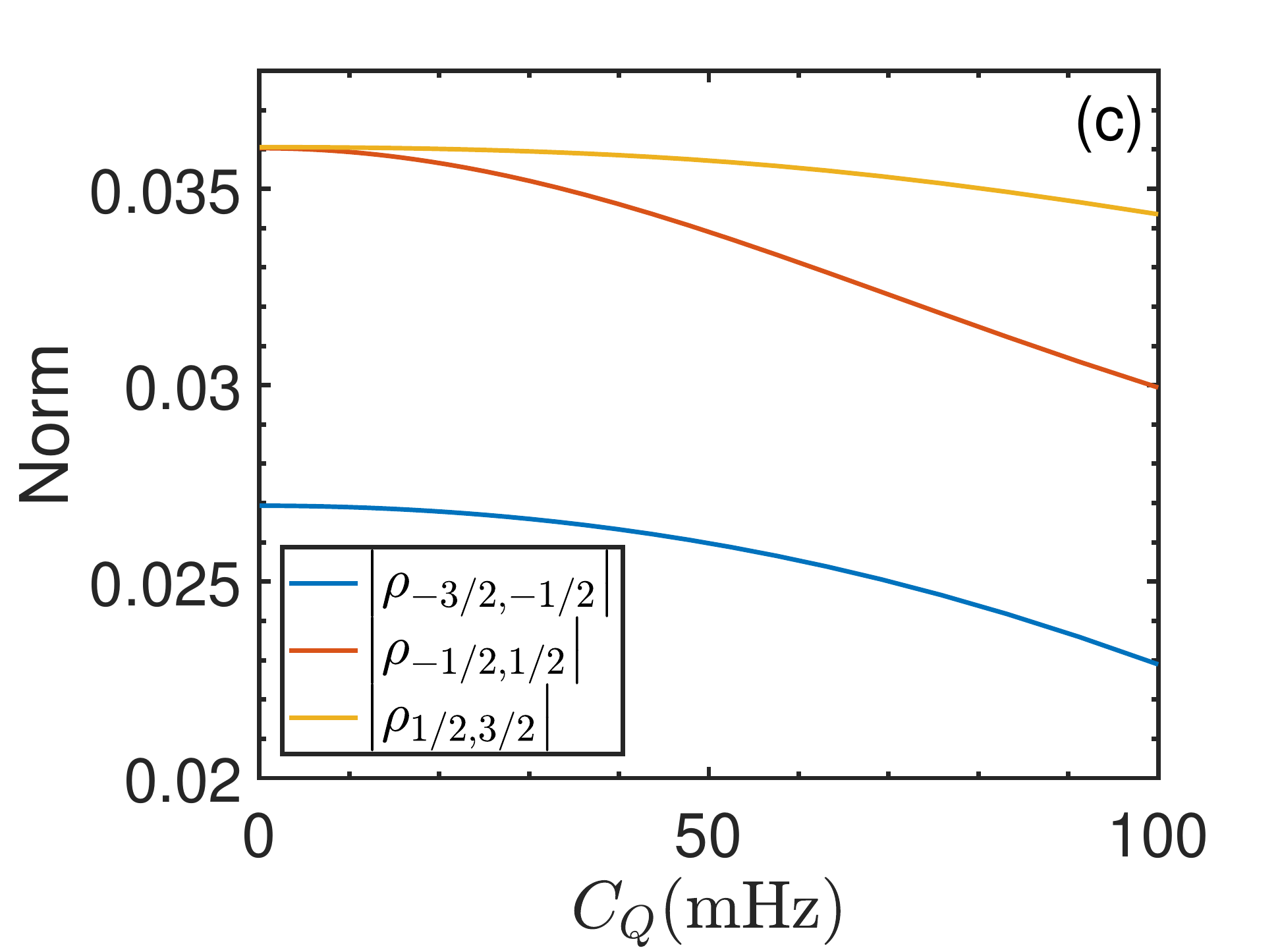} %
\includegraphics[width=0.48\linewidth]{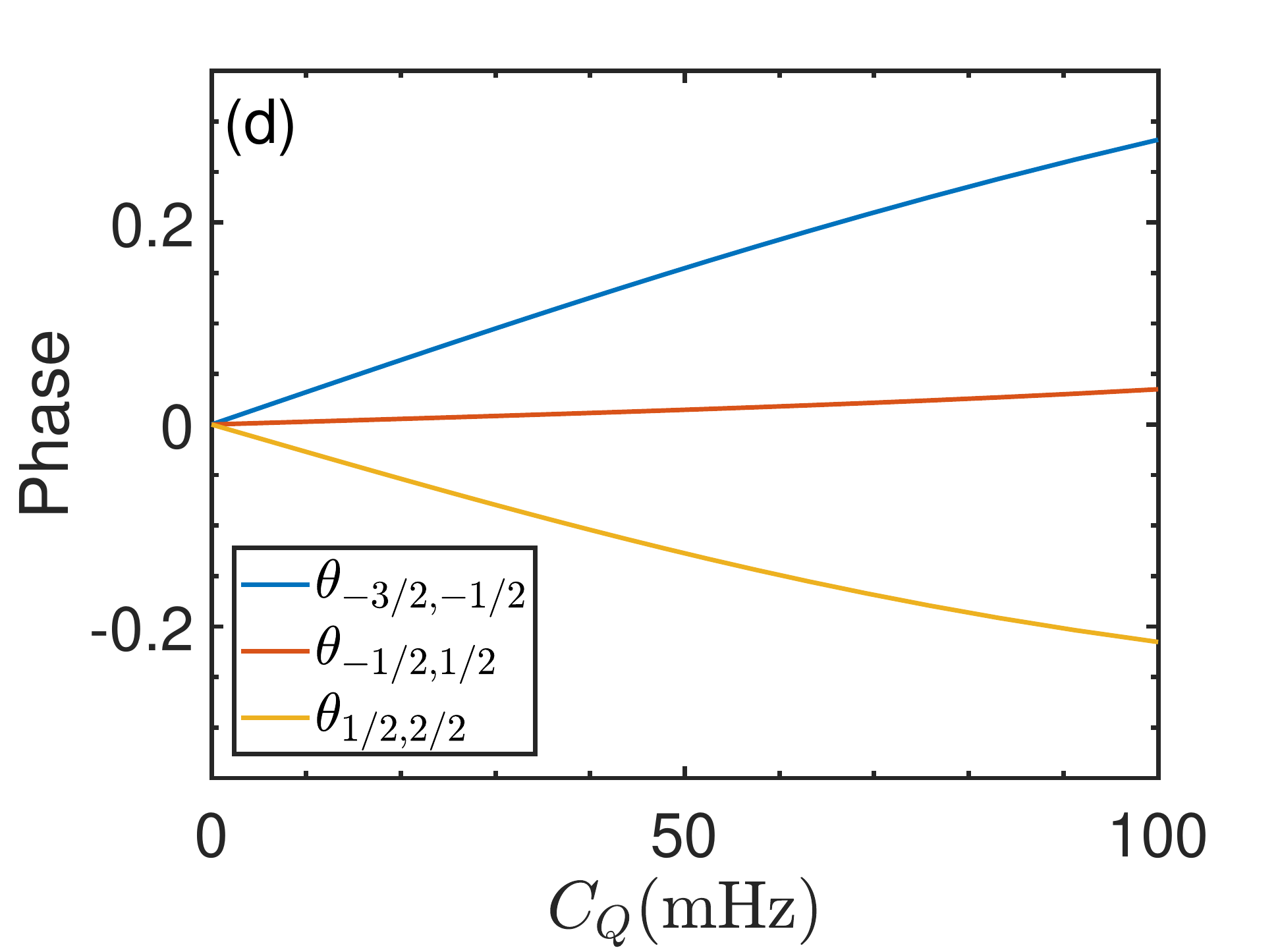}
\end{center}
\caption{(a) Precession frequency deviation $\protect\omega _{s}$ and (b)
average value $k_{x}$ as functions of the quadrupole interaction constant $%
C_{Q}$ at different amplitudes $\Omega _{d}$ of the drive. The legend in (b)
is the same as (a). It shows that, when $C_{Q}$ increases, the norm of the
frequency deviation $\protect\omega _{s}$ increase, while $k_{x}$, the
precession amplitude decrease. The later is resulted from the decrease of
the norm $\left\vert \protect\rho _{m,m+1}\right\vert $ and the difference
of the phase $\protect\theta _{m,m+1}$ of $\protect\rho _{m,m+1}$, which are
shown in (c) and (d) with $\Omega _{d}=0.2$Hz.}
\label{fig3}
\end{figure}

The dependence of the deviation $\omega _{s}$ and the average value $k_{x}$
on the quadrupole interaction constant $C_{Q}$ is shown in Figs.~\ref{fig3}%
(a) and (b) with different driving amplitudes $\Omega _{d}$, while other
parameters are the same as in Fig.~\ref{fig2}. When $C_{Q}$ increases, the
actual precession frequency deviates more from $\omega _{0}$. This brings
inaccuracy to the measurement on the precession frequency, because the exact
quadrupole interaction strength $C_{Q}$ is difficult to measure. Meanwhile,
as shown in Fig.~\ref{fig3}(b), $k_{x}$, which is $K_{\perp }$, the
precession amplitude in the original frame, decreases. Noting that in the
basis $\left\vert m\right\rangle $ ($m=-K,...,K$), which is the eigenstate
of $K_{z}$: $K_{z}\left\vert m\right\rangle =m\left\vert m\right\rangle $, $%
K_{x}$ is written as%
\begin{eqnarray}
K_{x} &=&\frac{\sqrt{3}}{2}\left\vert \frac{3}{2}\right\rangle \left\langle
\frac{1}{2}\right\vert +\left\vert \frac{1}{2}\right\rangle \left\langle -%
\frac{1}{2}\right\vert  \notag \\
&&+\frac{\sqrt{3}}{2}\left\vert -\frac{1}{2}\right\rangle \left\langle -%
\frac{3}{2}\right\vert +\text{H.c.},
\end{eqnarray}%
we plot the norm $\left\vert \rho _{m,m+1}\right\vert $ and phase $\theta
_{m,m+1}$\ of density matrix elements $\rho _{m,m+1}\equiv \left\langle
m\right\vert \tilde{\rho}_{s}\left\vert m+1\right\rangle $ where $%
m=-3/2,-1/2,$ and $1/2$, in Figs.~\ref{fig3}(c) and (d) with $\Omega
_{d}=0.2 $Hz while other parameters the same as in Figs.~\ref{fig3}(a) and
(b). Figs.~\ref{fig3}(c) and (d) show that the decrement of $k_{x}$ results
from two sources: the norm $\left\vert \rho _{m,m+1}\right\vert $ decreases
and the phase $\theta _{m,m+1}$ is nonzero. Both of these comes from the
destructive interference between oscillations (in the original frame)
between states $\left\vert m\right\rangle $ and $\left\vert m+1\right\rangle
$: because of the quadrupole interaction, the inherent energies of states $%
\left\vert m\right\rangle $ and $\left\vert m+1\right\rangle $ are different
(the eigenvalues of $\tilde{H}$ with $\Omega _{d}=0$). Thus, in presence of
the dissipation and feedback drive, in the steady state, all these
oscillation frequencies become $\omega _{0}-\omega _{s}$, but with different
phases of $\rho _{m,m+1}$, which leads to destructive interference in the
amplitude $K_{\perp }$.

\begin{figure}[tbp]
\begin{center}
\includegraphics[width=0.48\linewidth]{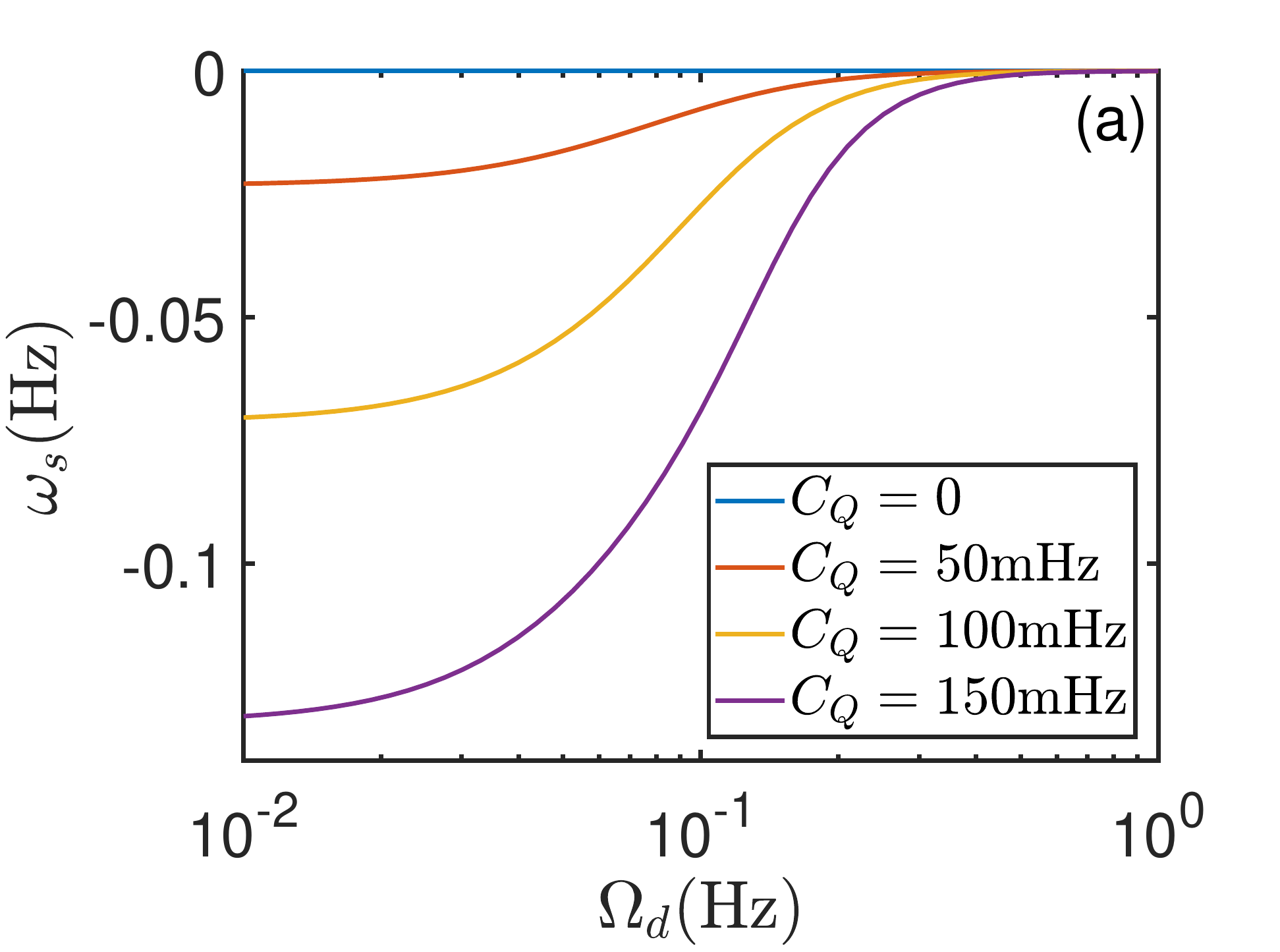} %
\includegraphics[width=0.48\linewidth]{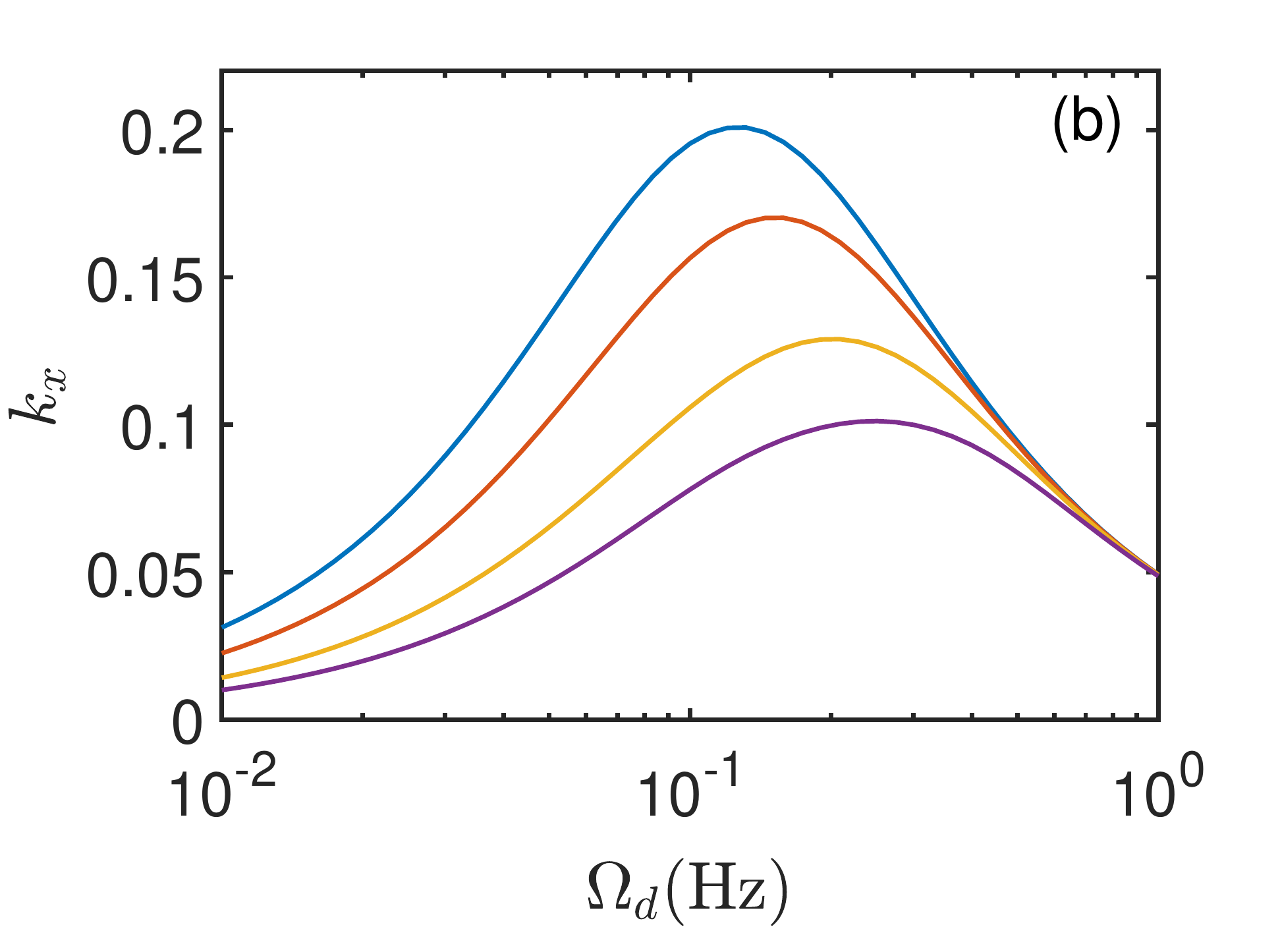}
\end{center}
\caption{(a) $\protect\omega _{s}$ as a function of the driving amplitude $%
\Omega _{d}$, for different quadrupole interaction constant $C_{Q}$. It
shows that $\left\vert \protect\omega _{s}\right\vert $ decreases
monotonically when $\Omega _{d}$ increases. At large $\Omega _{d}$, the
system behaves like a system without quadrupole interaction. This can be
seen from (b), where for different values of $C_{Q}$, $k_{x}$ converges to a
value in the $C_{Q}=0$ case. }
\label{fig4}
\end{figure}

The decrement of the precession amplitude can be compensated by increasing
the density of the noble gas, but the deviation of the precession frequency remains
problematic. By increasing the driving amplitude $\Omega _{d}$,\ we find
that the deviation of the precession frequency decreases, as shown in Fig.~%
\ref{fig4}(a), with other parameters the same as in Fig.~\ref{fig3}. Since
Eq.~(\ref{5}) gives that the solution $\omega _{s}$ is inverted when the the
quadrupole interaction constant $C_{Q}$ is inverted, we conclude that the
absolution value $\left\vert \omega _{s}\right\vert $ monotonically
decreases by increasing $\Omega _{d}$. Physically, this can be explained
intuitively: with larger $\Omega _{d}$, the nonlinear effect of the
quadrupole interaction is relatively smaller. Namely, the eigenvalues of the
Hamiltonian $\tilde{H}$ ($\beta =0$) is relatively more equally spaced. In
this case, the system is closer to a system without quadrupole interaction.
This can be shown in Fig.~\ref{fig4}(b), where at large $\Omega _{d}$\
limit, for different values of $\Omega _{d}$, $k_{x}$ with $C_{Q}>0$\
coincides with the curve with $C_{Q}=0$.

\begin{figure}[tbp]
\begin{center}
\includegraphics[width=0.48\linewidth]{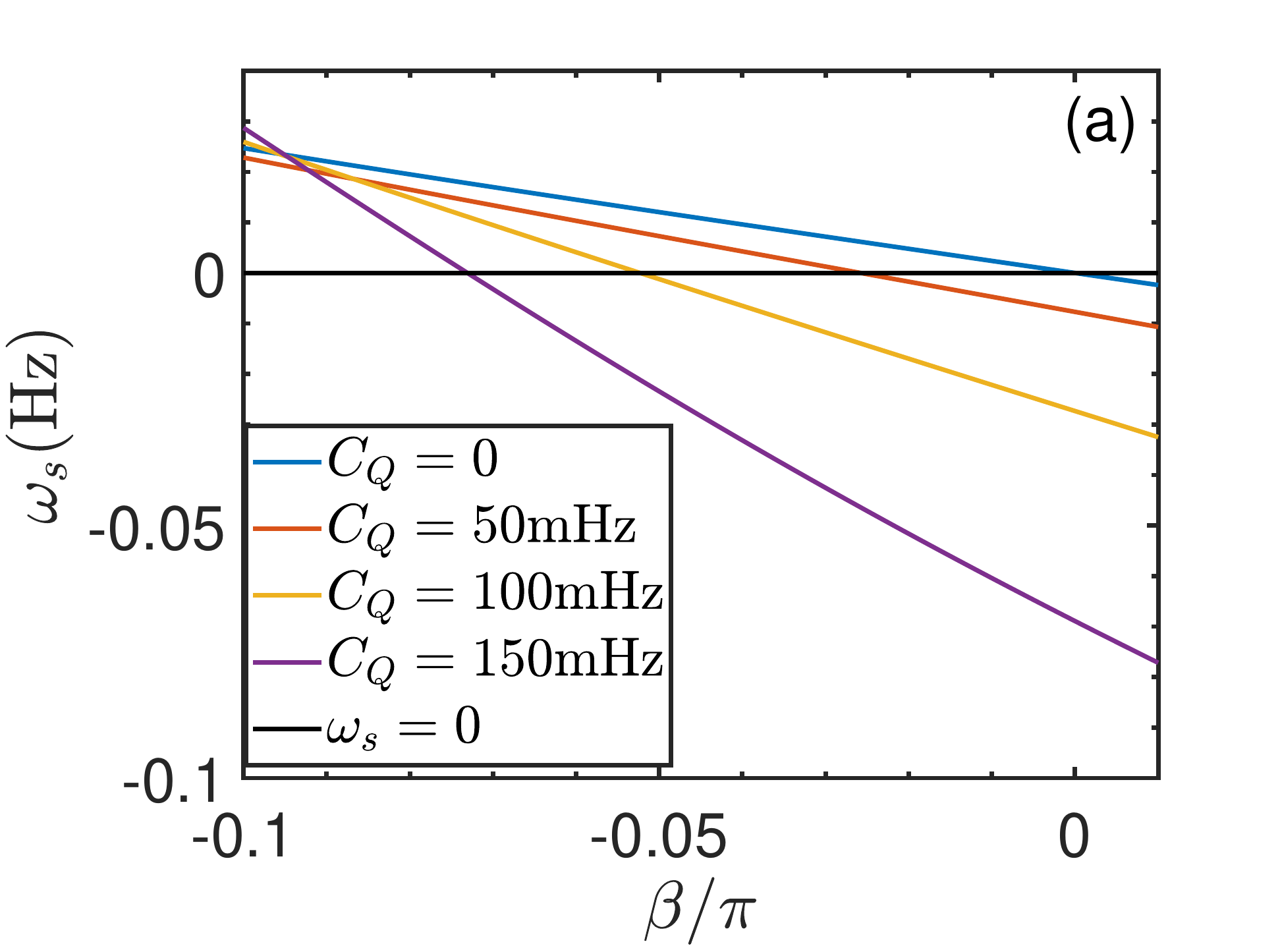} %
\includegraphics[width=0.48\linewidth]{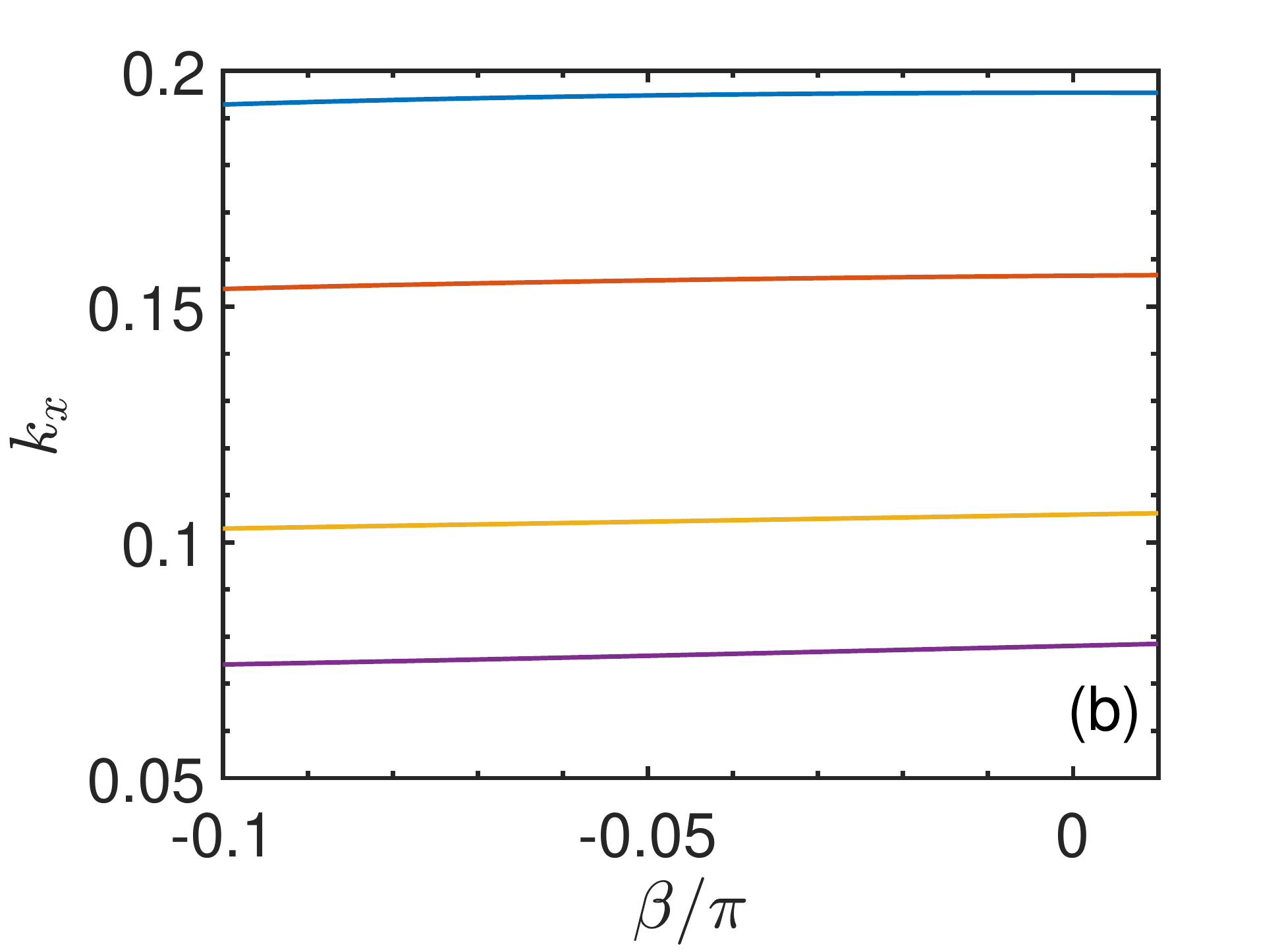}
\end{center}
\caption{(a) Precession frequency deviation $\protect\omega _{s}$ and (b)
precession amplitude $k_{x}$ as functions of the phase delay $\protect\beta $
for different values of $C_{Q}$. The black line in (a) has $\protect\omega %
_{s}=0$, and its intersection with the curves gives a value $\protect\beta %
_{0}$ that makes $\protect\omega _{s}=0$. This value $\protect\beta _{0}$
varies with different $C_{Q}$. Meanwhile, (b) shows that as the phase delay $%
\protect\beta $ changes, the precession amplitude $k_{x}$ almost remains
unchanged.}
\label{fig5}
\end{figure}

Another way to reduce the quadrupole shift is by tuning the phase delay $%
\beta $ \cite{walker2016spin}. The change of $\omega _{s}$ as a function of $%
\beta $ is shown in Fig.~\ref{fig5}a, where $\Omega _{d}=0.1$Hz and other
parameters the same as in Fig.~\ref{fig4}. Here, the intersection with the
black line gives a value $\beta _{0}$ that makes $\omega _{s}=0$. Around $%
\beta _{0}$, $\left\vert \omega _{s}\right\vert $ is smaller than in the $%
\beta =0$ case. However, $\beta _{0}$ is strongly depending on the system
parameters. For instance, when $C_{Q}$ becomes $-C_{Q}$, $\beta _{0}$
changes to $-\beta _{0}$ (Eq.~(\ref{5})). As a result, to reduce $\left\vert
\omega _{s}\right\vert $ by tuning the phase delay $\beta $,\ one would need
the knowledge $C_{Q}$, which makes it less practical than simply increasing
the driving amplitude $\Omega _{d}$. Moreover, tuning $\beta $ does not help
to increase the precession amplitude, as show in Fig.~\ref{fig5}b.
Therefore, in the following, we only focus on the case with $\beta =0$.

So far, we have studied the precession frequency in a relatively small
quadrupole interaction regime, where only one precession frequency exists.
In the next section, we will show that when $C_{Q}$ becomes large,
more precession frequencies appear.

\section{Large $C_{Q}$ regime: multi-precession-frequencies}

\label{s5}

For small quadrupole interaction constant $C_{Q}$, in the long-term limit,
only one precession frequency exists, and increasing the driving amplitude
can monotonically decrease the frequency deviation. However, this is not the
case for large $C_{Q}$, where the quadrupole splitting ($2C_{Q}$ for
spin-3/2) is not much smaller than the characteristic decay rate of the
system that is determined by the relaxation rates $\Gamma _{1}$, $\Gamma
_{2} $, $\Gamma _{p}$, and the driving amplitude $\Omega _{d}$. In this
regime, more than one precession frequency appears and the long-term behavior
of the system depends on initial conditions of the system.

\begin{figure}[tbp]
\begin{center}
\includegraphics[width=0.48\linewidth]{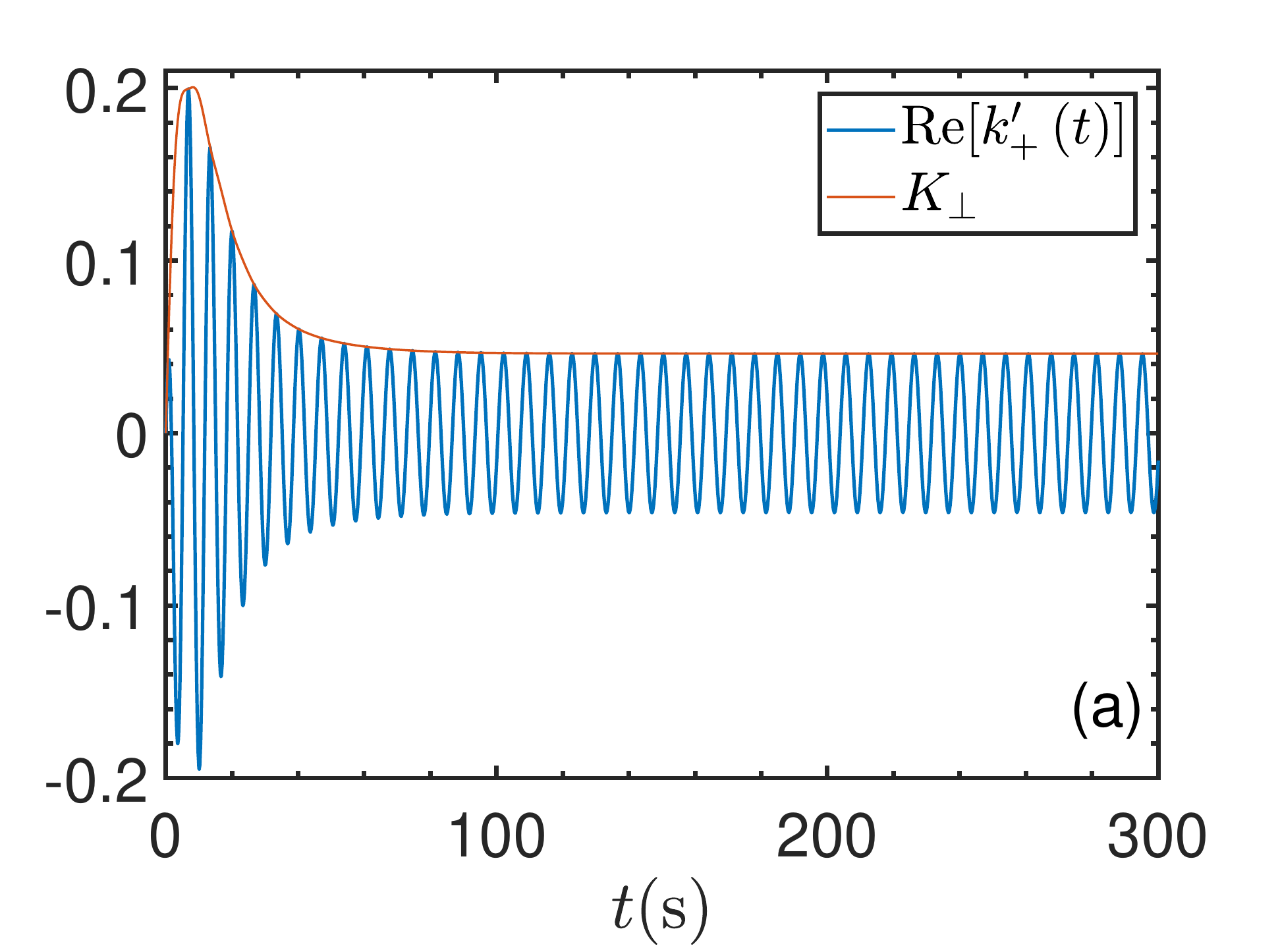} %
\includegraphics[width=0.48\linewidth]{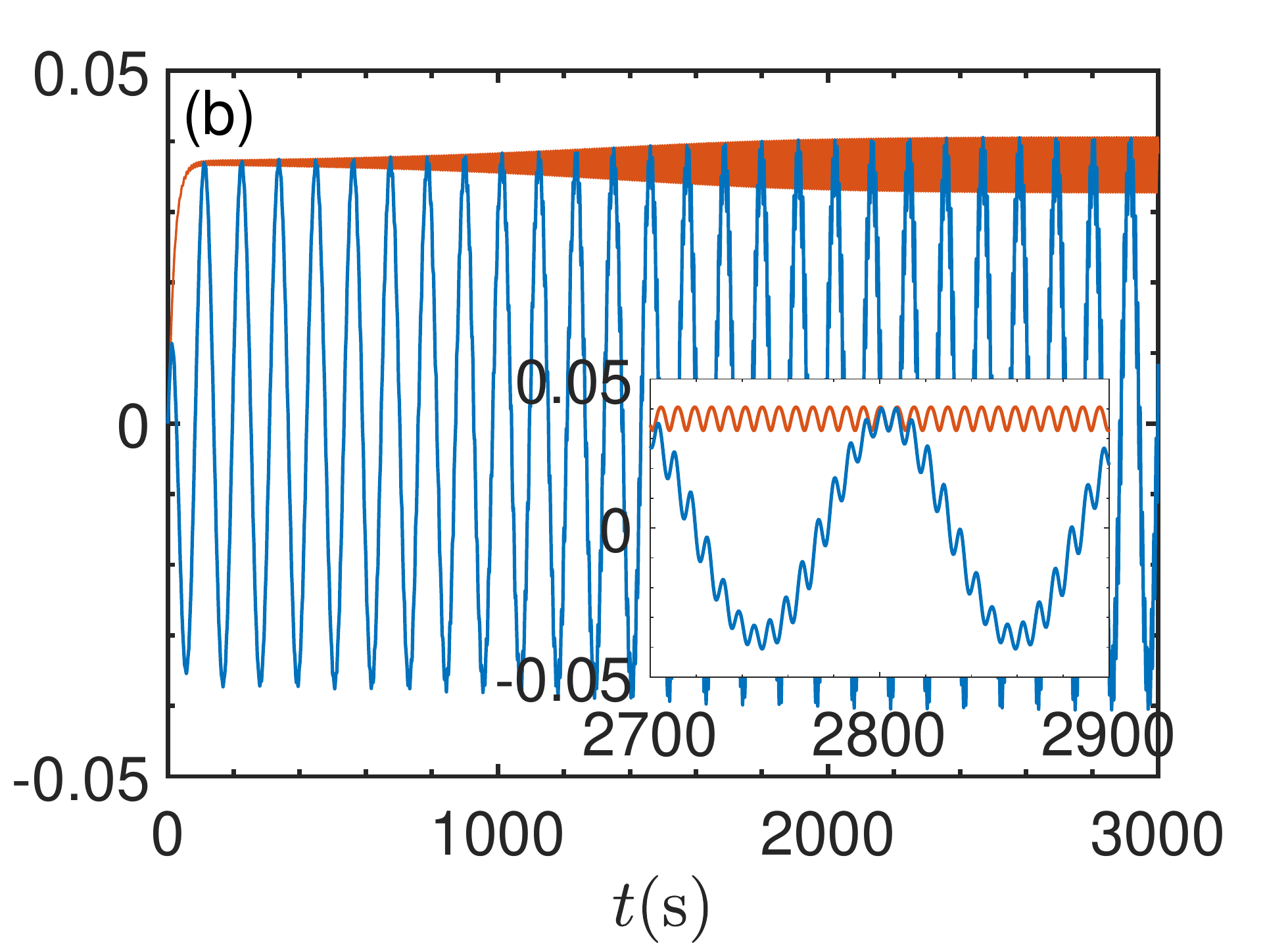} %
\includegraphics[width=0.48\linewidth]{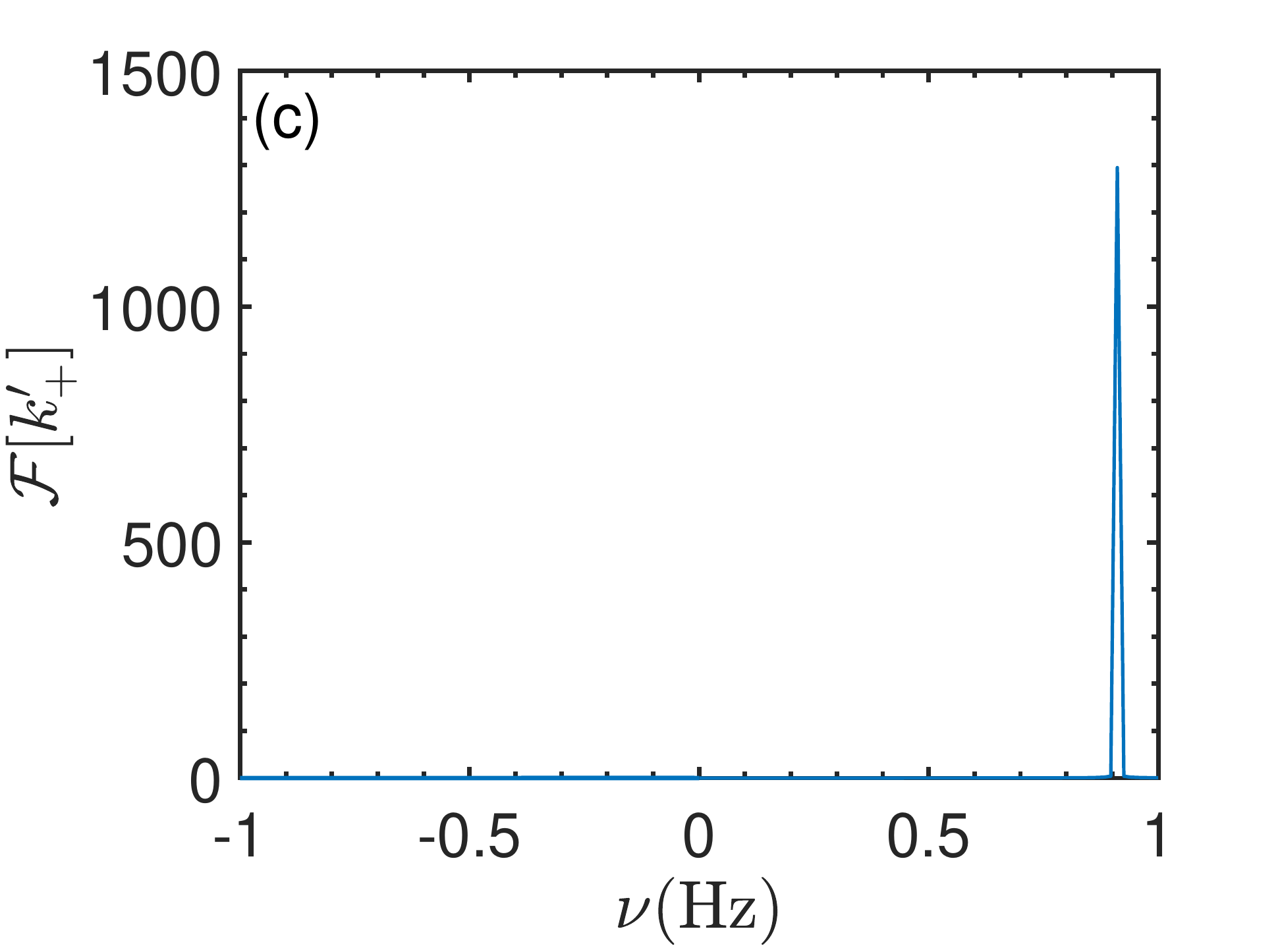} %
\includegraphics[width=0.48\linewidth]{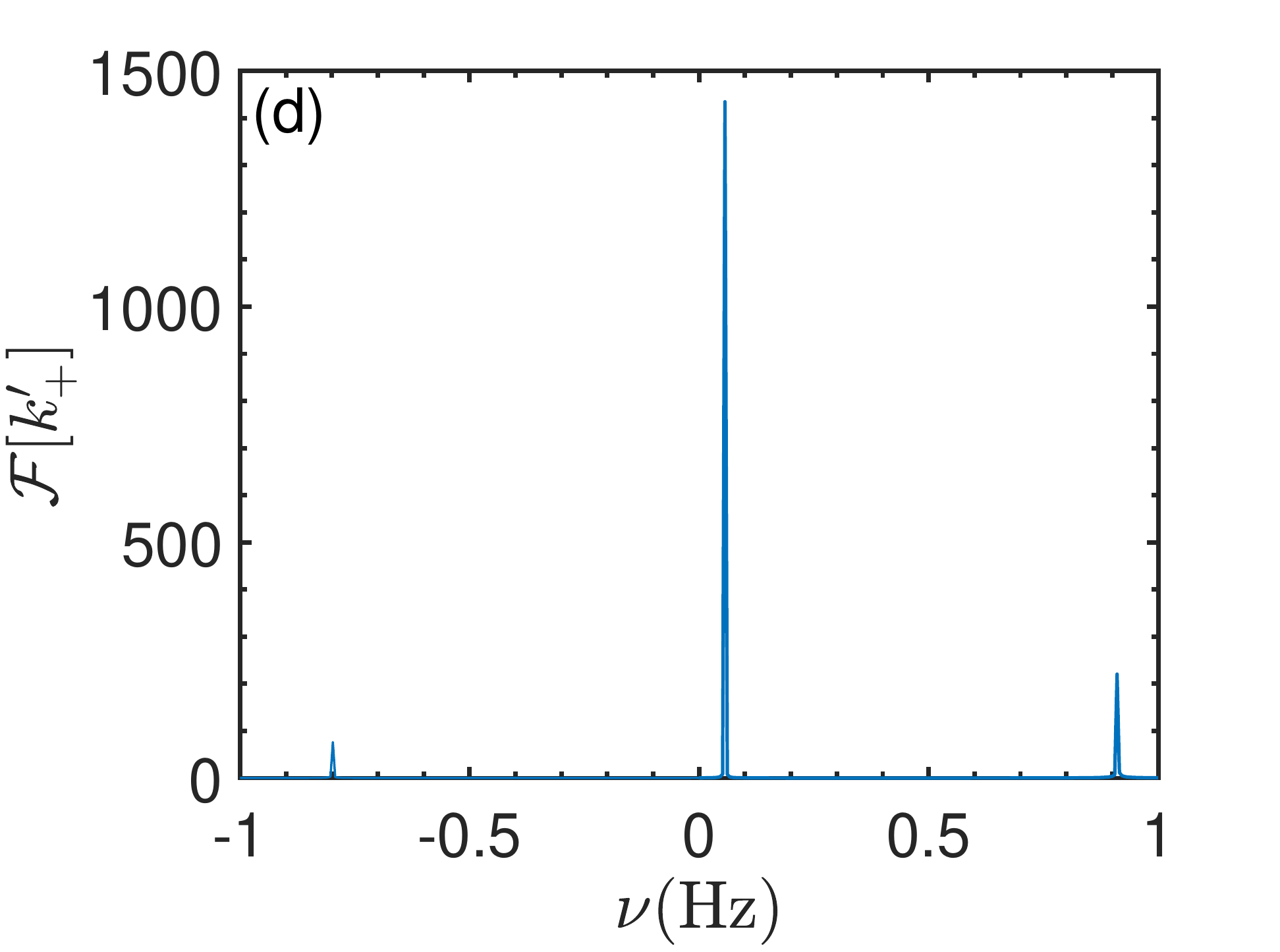}
\end{center}
\caption{Nuclear spin polarization along the $x$ direction Re$\left[
k_{+}^{\prime }\left( t\right) \right] $ (in the rotating frame with respect
to $\protect\omega _{0}K_{z}$) for different initial states (a) $\tilde{%
\protect\rho}_{0}^{\prime }=\left\vert 3/2\right\rangle \left\langle
3/2\right\vert $ and (b) $\tilde{\protect\rho}_{0}^{\prime }=I$. The Fourier
transform of $k_{+}^{\prime }\left( t\right) $ in the long-term limit from
the time domain to frequency domain is shown in (c) and (d), with the
corresponding initial condition the same as (a) and (b), respectively. Both
the behaviors in time and frequency domains show that the nuclear spins may
precess with a single frequency or multi-frequencies, depending on the
initial states. One of the frequencies in the latter case is the same as the
single frequency in (c), and these multi-frequencies are equally spaces, as
shown in (d).}
\label{fig6}
\end{figure}

To separate the fast oscillation with frequency $\omega _{0}$, we work in a
rotating frame with respect to $\omega _{0}K_{z}$, where the dynamics of the
density matrix $\tilde{\rho}^{\prime }$ is governed by Eq.~\ref{3} with $%
\omega =0$. In this frame, the average value $k_{+}^{\prime }\equiv \mathrm{%
Tr}\left( K_{+}\tilde{\rho}^{\prime }\right) $ is connected to $\left\langle
K_{+}\right\rangle $ in the original frame as%
\begin{equation}
\left\langle K_{+}\right\rangle =k_{+}^{\prime }e^{i\omega _{0}t},
\end{equation}%
and $\left\vert k_{+}^{\prime }\right\vert =K_{\perp }$. To show how the
long-term behavior depends on initial conditions, we plot Re$\left[
k_{+}^{\prime }\left( t\right) \right] $ and $K_{\perp }$ for two different
initial states $\tilde{\rho}_{0}^{\prime }$: (a) $\tilde{\rho}_{0}^{\prime
}=\left\vert 3/2\right\rangle \left\langle 3/2\right\vert $ and (b) $\tilde{\rho%
}_{0}^{\prime }=I$ in Figs.~\ref{fig6}(a) and (b), with $\Omega _{d}=0.1$Hz.
In case (a), Re$\left[ k_{+}^{\prime }\left( t\right) \right] $ oscillates
with a single frequency and $K_{\perp }$ is a constant in large-$t$ limit.
However, in Figs.~\ref{fig6}(b), Re$\left[ k_{+}^{\prime }\left( t\right) %
\right] $ shows multi-frequencies and its amplitude $K_{\perp }$ also has
oscillations.$\ $The Fourier transform of $\mathcal{F}[k_{+}^{\prime }]$
gives the precession frequencies, as shown in the peaks in Figs.~\ref{fig6}%
(c) and (d) corresponding to cases (a) and (b), respectively. It is shown
that besides a frequency $\upsilon _{0}=0.91$Hz that is the same as in case
(a), $k_{+}^{\prime }$ has two more frequencies $\upsilon _{1}=0.056$Hz and $%
\upsilon _{-1}=-0.80$Hz in case (b). The weights of oscillations with these
three frequencies are different, but the differences between the adjacent
frequencies are the same (considering the numerical error): $\upsilon
_{1}-\upsilon _{0}=\upsilon _{0}-\upsilon _{1}$. The reason is because in
the Hamiltonian $H$ without drive, the quadrupole interaction $%
C_{Q}K_{z}^{2} $ causes two more frequencies in oscillations between
states $\left\vert 3/2\right\rangle $ and $\left\vert 1/2\right\rangle $ and
 states $\left\vert -1/2\right\rangle $ and $\left\vert -3/2\right\rangle $, and the
spacing between these two frequencies and the ideal precession frequency $%
\omega _{0} $ are equal. Therefore, under the feedback drive and relaxation,
$k_{+}^{\prime }\left( t\right) $ oscillates with equally spaced frequencies.

\begin{figure}[tbp]
\begin{center}
\includegraphics[width=0.48\linewidth]{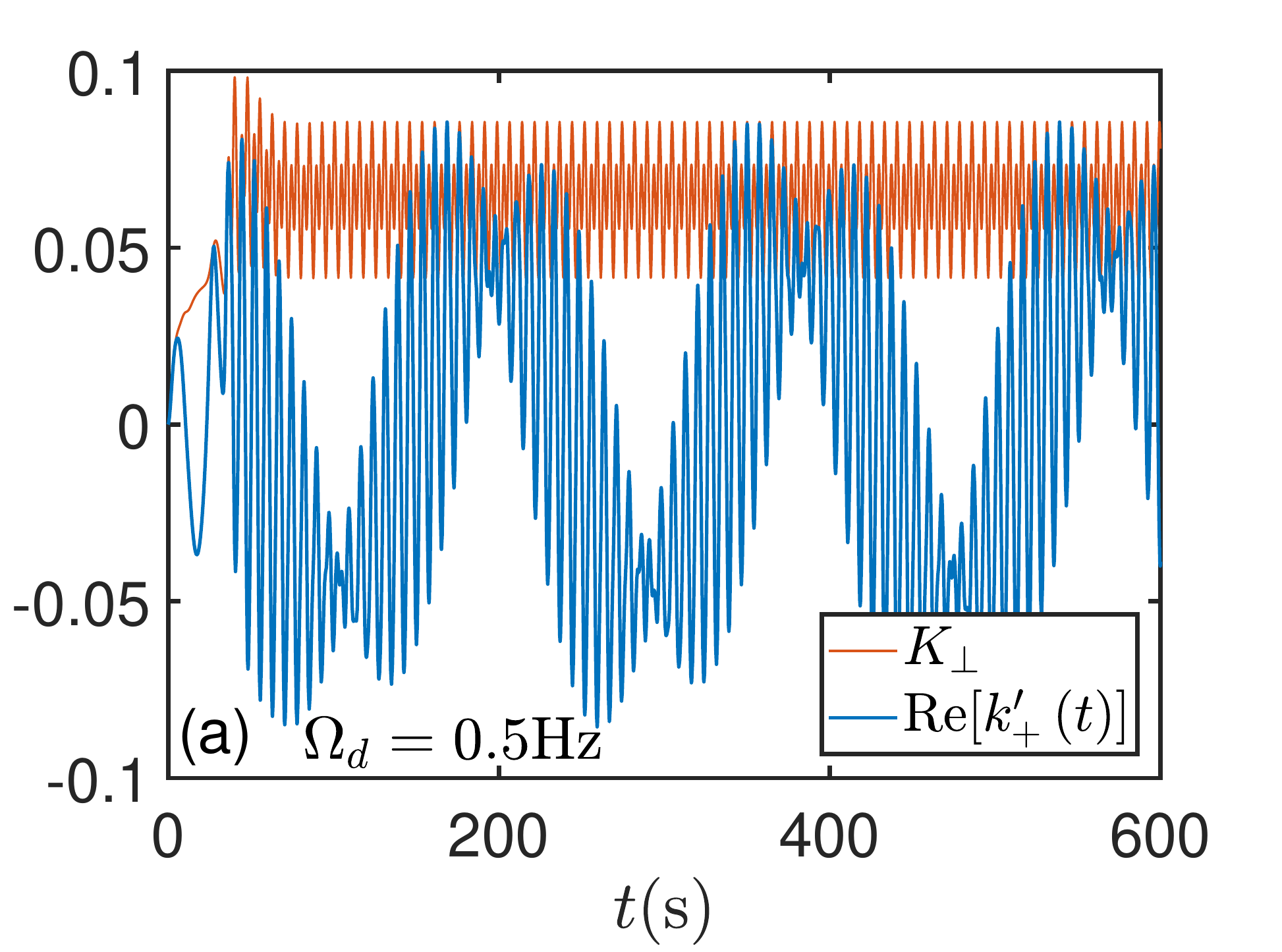} %
\includegraphics[width=0.48\linewidth]{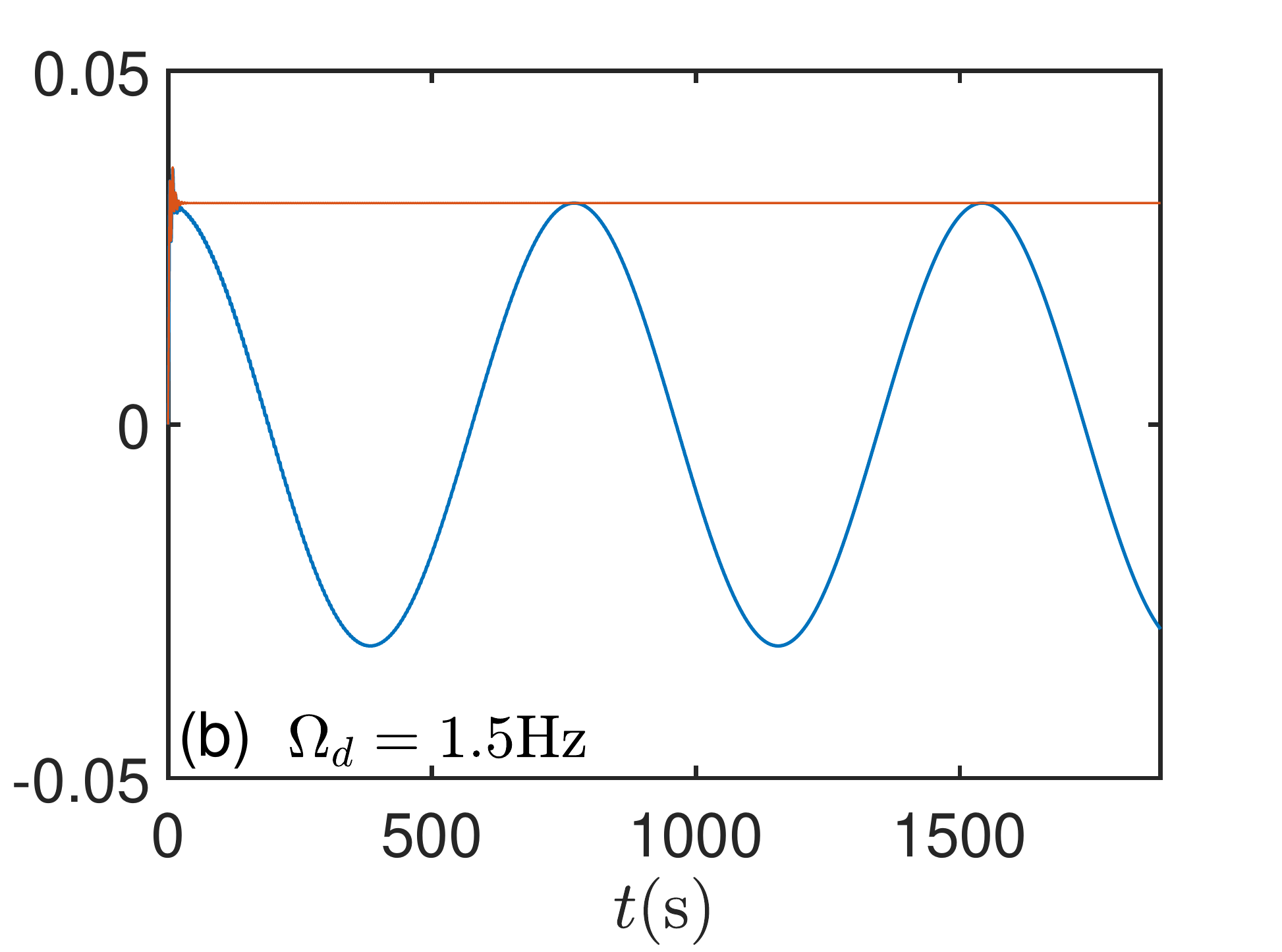} %
\includegraphics[width=0.48\linewidth]{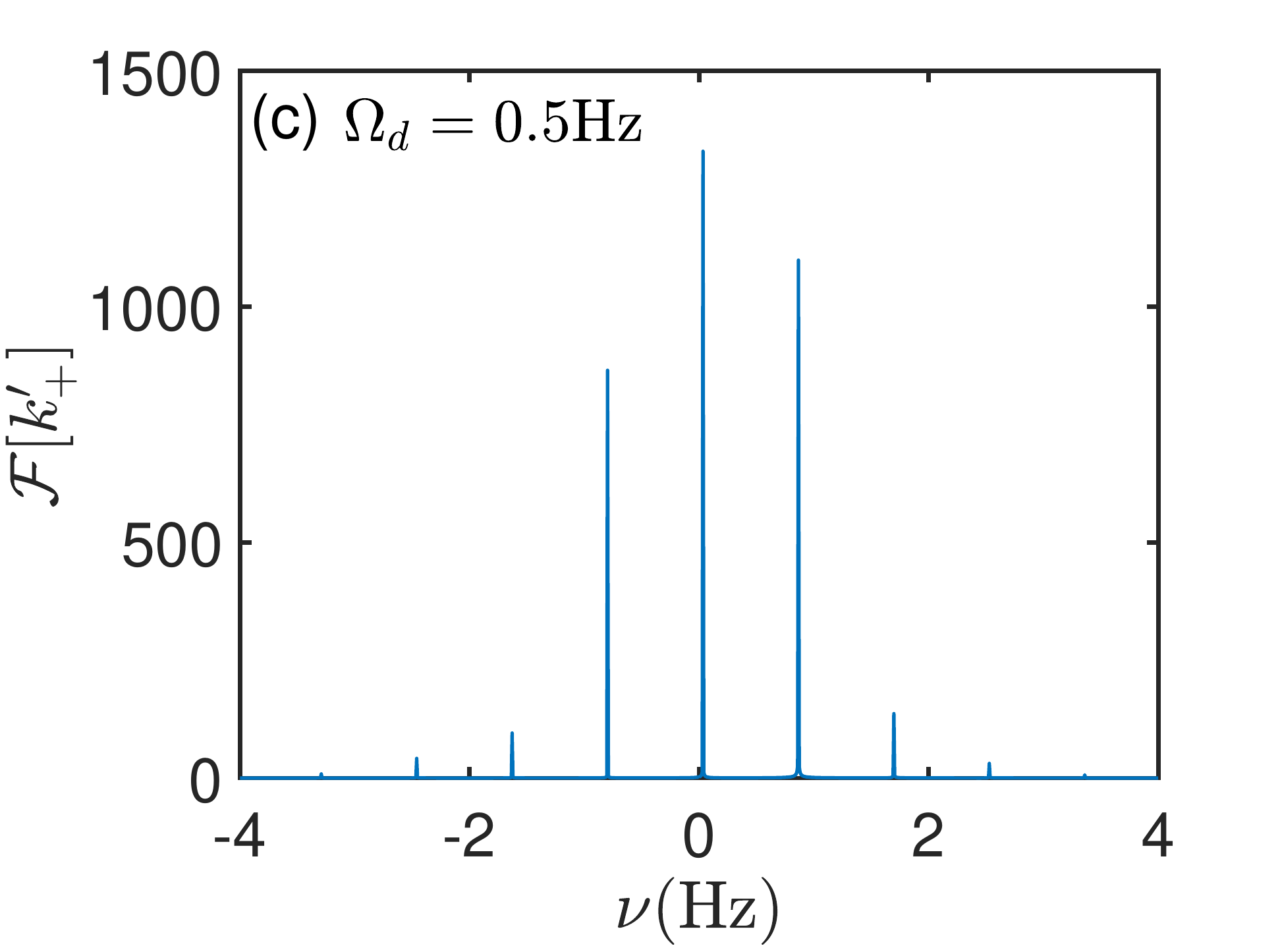} %
\includegraphics[width=0.48\linewidth]{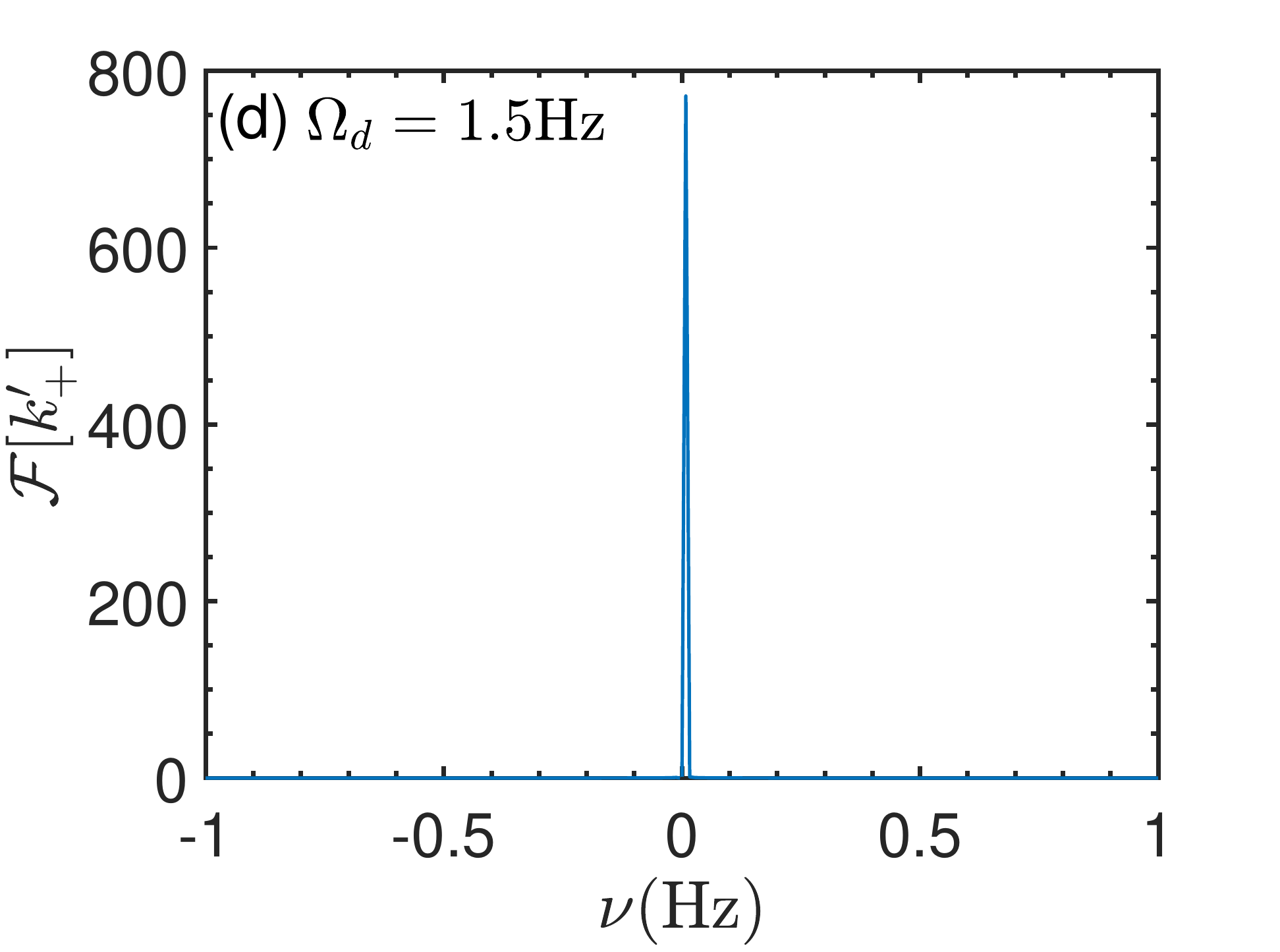}
\end{center}
\caption{Same as Fig.~\protect\ref{fig6} for $\tilde{\protect\rho}%
_{0}^{\prime }=I$, but with different driving amplitudes $\Omega _{d}=0.5$Hz
for (a) and (c), $\Omega _{d}=1.5$Hz for (b) and (d). Figures (a) and (c)
show that increasing \ $\Omega _{d}$ will increase the number of the
precession frequencies. These frequencies are equally spaces. However, the
further increment of $\Omega _{d}$ can drive the nuclear spin to precess
with a single frequency, as shown in (b) and (d).}
\label{fig7}
\end{figure}

When the driving amplitude $\Omega _{d}$ increases to $0.5$Hz, more
oscillations are stimulated, as shown in Fig.~\ref{fig7}, where the
precession amplitue $K_{\perp }$ and the nuclear spin's polarization in the $%
x$ direction Re$\left[ k_{+}^{\prime }\left( t\right) \right] $ are plotted
in (a), while the Fourier transform $\mathcal{F}[k_{+}^{\prime }]$ is
plotted in (c). Here, the initial state $\tilde{\rho}_{0}^{\prime }=I$.\
Similar to the case in Fig.~\ref{fig6}(d), the nine oscillations have
different weights while the frequencies are equally spaced. As $\Omega _{d}$
is further increased, the system comes back to the small quadrupole
interaction case: the nuclear spin precesses with a single frequency, as
shown in Fig.~\ref{fig7}(b) and (d), with $\Omega _{d}=1.5$Hz. Meanwhile,
the frequency deviation ($0.008$Hz) is smaller than the smallest frequency
deviation in cases with smaller $\Omega _{d}$ ($0.056$Hz for $\Omega
_{d}=0.1 $Hz and $0.034$Hz for $\Omega _{d}=0.5$Hz). The reason is the same
as explained in Sec.~\ref{s5}: the nonlinearity resulting from the
quadrupole interaction is reduced by very large $\Omega _{d}$.

\section{Conclusion}

\label{s6} We have theoretically studied the effects of quadrupole
interactions on feedback-generated-driven NMRGs whose nuclear spins are
larger than $1/2$. This quadrupole interaction that is resulting from
collisions of the noble gas atoms with the cell walls, causes a single shift in the
nuclear spin's precession frequency when the quadrupole interaction constant
$C_{Q}$ is small compared to the characteristic decay rate of the system.
Solving the effective master equation of the nuclear spins by neglecting its
correlations with the electron spins, we show that the quadrupole shift can
be monotonically decreased by increasing the amplitude of the feedback drive.

In large $C_{Q}$ regime, the nuclear spin may precess with a single
frequency or multi-frequencies depending on its initial state, since the
master equation is nonlinear. We have studied the dynamics of the system in
the long-term limit. It is shown that when the driving amplitude is larger
enough, the nuclear spin restores the single-frequency precession.

\acknowledgments This work was supported by the National Key R$\&$D Program of China Grant No.2017YFB0503102 and the National Natural Science
Foundation of China Grant No.61627806.

\bibliography{v3.bbl}

\end{document}